\documentclass[twocolumn,english,prl,superscriptaddress]{revtex4-1}
\usepackage[utf8]{inputenc}
\usepackage{color}
\usepackage{babel}
\usepackage{bm}
\usepackage{graphicx}
\usepackage{physics}
\usepackage{amsmath}
\usepackage{amssymb}
\usepackage{empheq}
\usepackage{times}

\usepackage[colorlinks=true,citecolor=blue]{hyperref}
\newcommand{\sect}[1]{\emph{{#1}}.---}

\begin{document}

\title{Unsupervised machine learning of quantum phase transitions using diffusion maps}

\author{Alexander Lidiak}
\email{alidiak@mines.edu}
\affiliation{Department of Physics, Colorado School of Mines, Golden, Colorado 80401, USA}
\author{Zhexuan Gong}
\email{gong@mines.edu}
\affiliation{Department of Physics, Colorado School of Mines, Golden, Colorado 80401, USA}
\affiliation{National Institute of Standards and Technology, Boulder, Colorado 80305, USA}

\date{\today}

\begin{abstract}

Experimental quantum simulators have become large and complex enough that discovering new physics from the huge amount of measurement data can be quite challenging, especially when little theoretical understanding of the simulated model is available. Unsupervised machine learning methods are particularly promising in overcoming this challenge. For the specific task of learning quantum phase transitions, unsupervised machine learning methods have primarily been developed for phase transitions characterized by simple order parameters, typically linear in the measured observables. However, such methods often fail for more complicated phase transitions, such as those involving incommensurate phases, valence-bond solids, topological order, and many-body localization. We show that the diffusion map method, which performs nonlinear dimensionality reduction and spectral clustering of the measurement data, has significant potential for learning such complex phase transitions unsupervised. This method may work for measurements of local observables in a single basis and is thus readily applicable to many experimental quantum simulators as a versatile tool for learning various quantum phases and phase transitions.

\end{abstract}

\maketitle

With the recent demonstration of quantum supremacy \cite{arute_quantum_2019}, the need for understanding well-controlled experimental quantum systems that cannot be simulated efficiently on a classical computer is growing rapidly. However, experimental data sets generated by measurements on post quantum-supremacy devices can be too large and complex for traditional data analysis tools to extract useful features from. This in particular poses a major challenge in using quantum simulators to make new discoveries at the frontier of quantum many-body physics, where existing theoretical understanding is often lacking \cite{johnson_what_2014}. A promising method to address this challenge is unsupervised machine learning, which can extract important features from data with little to no a priori understanding of the data \cite{VanNieuwenburg2017,Torlai2016, carleo2019machine,Broecker2017, Carrasquilla2017,broecker2017quantum,LeiWang2016,Ch_ng_2017,MLearning_new_phys,Greplova2020,torlai2019integrating,torlai2018neural}.

While machine learning has become a standard toolbox for data analysis in many areas of physics, including high-energy, astrophysics \cite{carleo2019machine}, and condensed-matter physics \cite{Zhang_2019}, the use of unsupervised machine learning in experimental quantum simulators, in particular for studying quantum phases and phase transitions, is so far lacking. For example, the standard approach to demonstrate a quantum phase transition in quantum simulation is to extract some feature from the measurement data related to an ``order parameter" \cite{bloch_quantum_2012,islam_onset_2011,zhang_observation_2017}. However, for the discovery of a new quantum phase or phase transition, it is often unclear what feature or order parameter one should extract from the data. For a simple symmetry breaking phase transition, one can usually find an order parameter that is linear in the measured observables. In such a scenario, a common unsupervised machine learning method known as principal component analysis (PCA) can be applied \cite{LeiWang2016, Hu2017, CeWang2017}, which performs a linear projection of the sample data onto a lower dimensional subspace for feature extraction. But for quantum systems with phases whose order parameters are complex, nonlinear functions of local observables \cite{carleo2019machine}, where machine learning could be particularly useful in discovering new physics, PCA will fail. Example systems include valence-bond solids \cite{Zhitomirsky_1996}, quantum spin liquids \cite{balents_spin_2010}, topologically ordered matter \cite{Wen:2017aa}, and many-body localized (MBL) systems \cite{nandkishore2015many}. A number of nonlinear dimensionality reduction methods in machine learning can be applied for these systems, such as kernel PCA \cite{CeWang2018}, auto-encoders \cite{Hu2017}, self-organizing maps \cite{Shirinyan2019}, and diffusion maps \cite{Rodriguez-Nieva2019}. The success so far is however limited, with a notable exception that the diffusion map method has recently been used to identify certain topological phases unsupervised \cite{Rodriguez-Nieva2019}, which has long been regarded as challenging. 

In this work, we show that the diffusion map is in fact a rather versatile method that can identify a variety of complex quantum phases. It is also computationally efficient and works for data easily obtained by quantum simulation experiments, such as the measurement of all spins in a single direction \cite{monroe2019programmable}. This is in contrast to many machine learning approaches \cite{VanNieuwenburg2017,Carrasquilla2017,Schindler_2017,Hsu2018,venderley2018machine,Matty_2019} that require the entanglement spectrum of quantum states, which is difficult to obtain experimentally. 

The main idea of the diffusion map is to reveal the structure of the measurement samples in configuration space and perform automatic clustering of the samples with a tunable cluster radius \cite{Coifman2006}. Major changes in the configurations of quantum states can be revealed as a result, suggesting the onset of a phase transition. As examples, we will show how diffusion maps can correctly identify incommensurate phases, valence-bond solid phases, and many-body localized phases, all of which are difficult to learn using traditional data analysis on expectation values, variances, and two-point correlations of local observables, or simple unsupervised machine learning methods such as PCA and k-means clustering \cite{supp}. Note that our primary goal is to learn phases of quantum matter from experimental data with minimal theoretical understanding. For precisely locating phase transition points, supervised learning methods can be applied afterwards \cite{VanNieuwenburg2017,Carrasquilla2017,Schindler_2017,Hsu2018}.

\sect{General picture} We will first introduce the diffusion map method and describe a general picture of how it can identify different quantum phases. In the following, the diffusion map is always applied to a collection of measurement samples. Without loss of generality, we assume each sample is from a measurement of $N$ quantum spins in some direction. Each measurement sample thus contains $N$ numbers, denoted by an $N$-dimensional vector $\bm{X}_i$ $(i=1,2,\cdots,M$). We obtain $M$ such samples by preparing and measuring the same state $M$ times, a routine practice in quantum simulation experiments \cite{monroe2019programmable}. One can also obtain these measurement samples computationally using either direct sampling if exact diagonalization is used \footnote{For exact diagonalization, we can obtain the ground or dynamical state as $|\psi\rangle = \sum_{S_k} \psi(S_k)|S_k\rangle$, where $|S_k\rangle$ is a basis state for the particular measurement and $k$ is a linearized index. We then pick a random number $0<x<1$ and if it falls in the window between $\sum_{i<k} \abs{\psi(S_i)}^2$ and $\sum_{i<k+1} \abs{\psi(S_i)}^2$ for some value of $k$, then the basis state $|S_k\rangle$ will be sampled.}, or Monte Carlo sampling if a variational ansatz is used \cite{schollwock_density-matrix_2011,Or_s_2014,carleo2017solving,dong-ling_deng_quantum_2017}.

The diffusion map sets a fictitious diffusion process among the samples based on their distances. First, a distance metric needs to be defined. Here, we use the normalized Euclidean distance between two samples $i$ and $j$, defined as $d_{ij}^2 \equiv \frac{1}{\mathcal N} \sum_{k=1}^N (X_{ik} -X_{jk})^2$, where $X_{ik}$ is the $k^{\text{th}}$ element of the sample vector $\bm{X}_i$, and $\mathcal N \sim N$ is a normalization constant that ensures $d_{ij}\in [0,1]$. Next, a kernel function is used to associate a transition probability between samples based on their distances. A common choice is the Gaussian kernel: $K_{ij}=e^{-d_{ij}^2/(2\epsilon)}$, where the hyperparameter $\epsilon$ controls how fast the transition probability decays with distance. Finally we introduce $P_{ij}=K_{ij}/(\sum_k K_{ik})$ as the normalized probability of the diffusion process from samples $i$ to $j$. 

Since the transition probability between any two samples is nonzero for a finite $\epsilon$, the above-mentioned diffusion process is ergodic in the long time limit. This means the largest eigenvalue of the $P$ matrix is always exactly $1$ \cite{Coifman2006,Rodriguez-Nieva2019}. If there are clusters of samples in which the samples have at most $r$ spins in different configurations, then up to a time scale $\tau = e^{r/N\epsilon}$, the diffusion process will be largely restricted within each cluster. The number of such clusters will correspond to eigenvalues of $P$ that are larger than $1-\delta$ where $\delta \sim 1/\tau$  \cite{Coifman2006}. Thus by choosing $\epsilon\sim r/(N\ln\tau)$, we can find the number of clusters of a particular size. For the examples we shall discuss, we keep the diffusion time $\tau$ fixed by choosing $\delta=10^{-2.5}$ (the exact value does not matter).

\begin{figure}
    \includegraphics[width=0.85\linewidth]{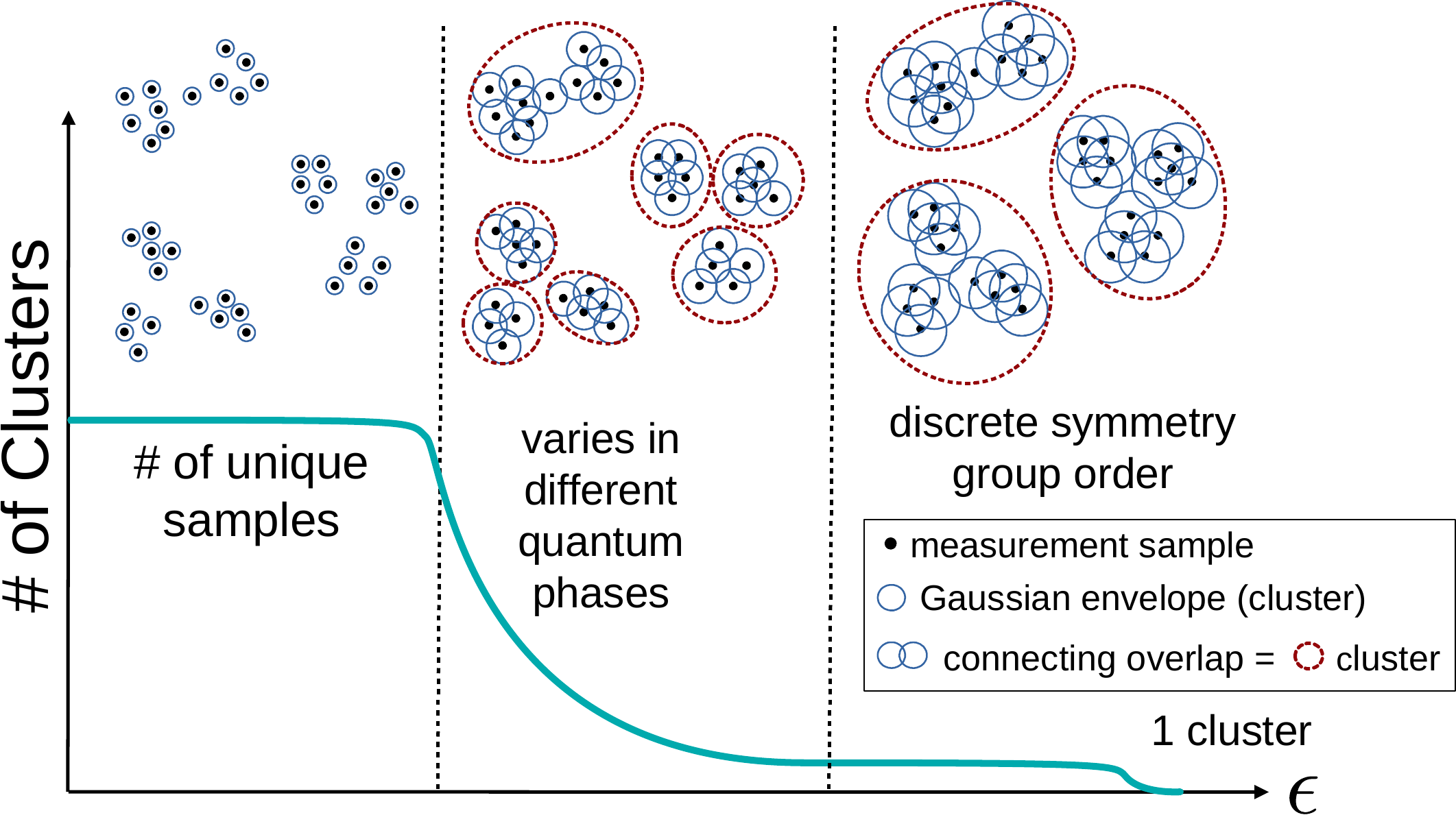}
	\caption{Schematic of how the diffusion map reveals the number of clusters formed by the measurement samples in configuration space. The hyperparameter $\epsilon$ determines the size of the clusters by controlling the width of the Gaussian envelope in the kernel function. The qualitative pictures for the small [$\epsilon\lesssim 1/(N\ln\tau)]$, intermediate, and large [$\epsilon\sim 1/\ln\tau$] $\epsilon$ regimes are shown. For sufficiently large $\epsilon$, the number of clusters becomes one.}
	\label{epsilon_regions}
\end{figure}

Schematically shown in Fig.\,\ref{epsilon_regions}, when $\epsilon\lesssim 1/(N\ln\tau)$, each different sample will be identified as one cluster, and the number of clusters in this regime is simply given by the number of unique measurement samples. For a generic quantum many-body state that contains non-negligible weights of exponentially many basis states, almost all measurement samples are different from each other. However, for a many-body localized quantum system, the number of unique samples can be significantly smaller than the number of samples. Thus a diffusion map with such a small $\epsilon$ should be able to distinguish an ergodic many-body state from a localized many-body state. On the other hand, when $\epsilon \sim 1/\ln\tau$, only samples with $\sim N$ spins in different configurations will belong to different clusters. If multiple clusters of samples are identified in this regime, one can expect a discrete spontaneous symmetry breaking in the measurement direction of the spins, with the order of the symmetry group equal to the number of clusters. Between the small and large $\epsilon$ regimes, we can tune $\epsilon$ to reveal the number of clusters with variable sizes (note that changing $\epsilon$ requires no additional experimental data). As shown below, the intermediate $\epsilon$ regime ($1/(N\ln\tau)\lesssim \epsilon \lesssim 1/\ln\tau$) is the key to learning a variety of complex quantum phases that cannot be identified using linear dimensionality reduction methods such as PCA. We emphasize that the diffusion map is computationally efficient for the number of samples ($10^2-10^3$) in a typical quantum simulation experiment \cite{monroe2019programmable}, and  we can compute diffusion maps with different system parameters or hyper-parameters in parallel.

\sect{Learning incommensurate phases} To demonstrate the power of diffusion maps with a tunable $\epsilon$ in learning complex quantum phases, we start with a $\mathbb{Z}_n$ transverse-field Ising model (TFIM) (also known as the chiral clock model \cite{Fendley2012}) with $n=3$. This model can be simulated using Rydberg atoms experimentally \cite{bernien_probing_2017,samajdar_numerical_2018} and has a nontrivial incommensurate phase between the usual ferromagnetic and paramagnetic phases of the TFIM. The Hamiltonian of the model reads $H_1 = -f \sum_{j=1}^{N} \tau_j e^{i\theta} - (1-f) \sum_{j=1}^{N-1} \sigma_j \sigma_{j+1}^{\dagger} e^{i\theta} +h.c.$, where $\sigma=\begin{psmallmatrix} 1 & 0 & 0 \\0 & e^{i 2\pi /3} & 0 \\0 & 0 & e^{-i 2\pi /3} \end{psmallmatrix}$ and $\tau=\begin{psmallmatrix} 0 & 0 & 1 \\ 1 & 0 & 0 \\0 & 1 & 0 \end{psmallmatrix}$ are the $\mathbb{Z}_3$ spin operators. 
Without chirality ($\theta=0$), the ground state of $H_1$ undergoes a simple  ferromagnetic (FM) to paramagnetic (PM) phase transition when increasing $f$ from $0$ to $1$. For $\theta>0$, an incommensurate (IC) phase appears for intermediate values of $f$, where spin correlations $\langle \sigma_i \sigma_j \rangle$ decay as a power law. The ferromagnetic phase can be easily identified using the average of $\{\langle\sigma_j\rangle\}$ which is nonzero in the FM phase. However, such an order parameter cannot tell the PM phase from the IC phase, as both phases have vanishing FM order. Therefore we find that PCA (as well as two-point correlations) cannot identify the IC phase and its boundary \cite{supp}.

\begin{figure}
    \includegraphics[width=\columnwidth]{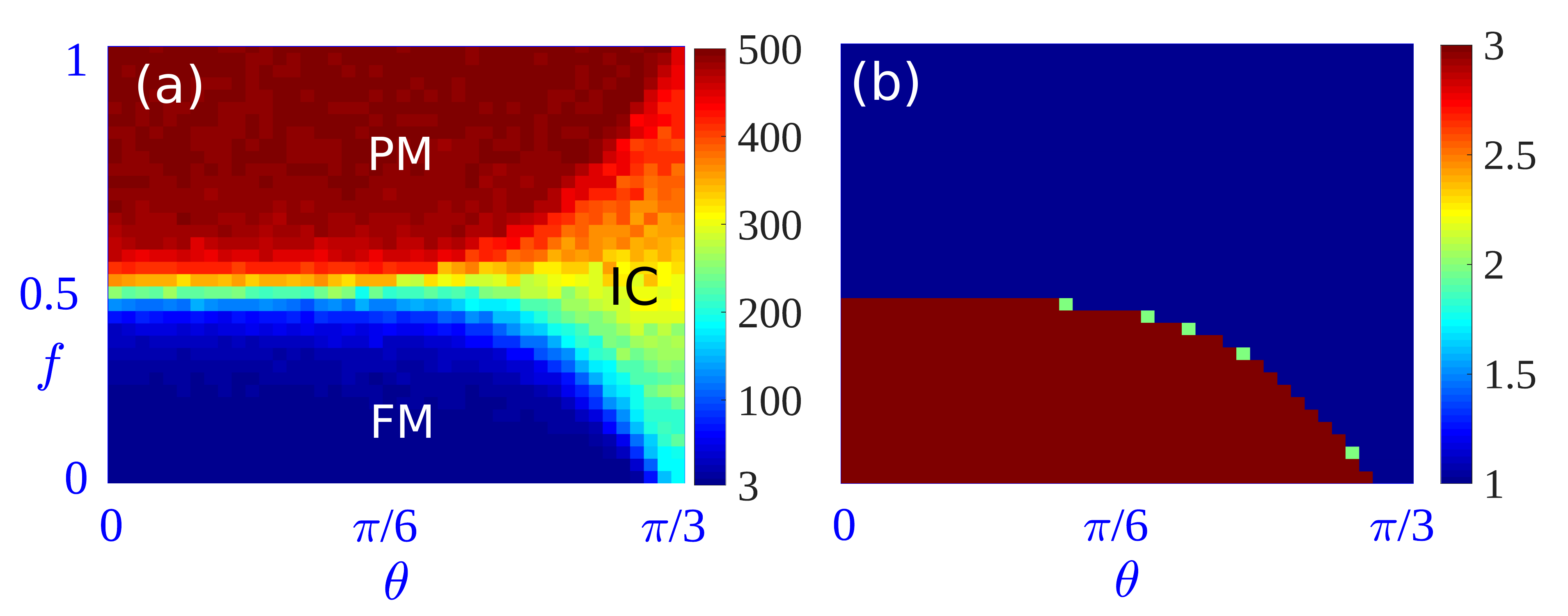} 
	\caption{ The ground-state phase diagram of the chiral $\mathbb{Z}_3$ TFIM (see $H_1$) found by the diffusion map method. The measurement samples are obtained from the ground-state found using exact diagonalization with $N=12$ spins. 500 samples are used for each value of $f$ and $\theta$. (a) $\epsilon=0.015$, an intermediate value that reveals all three phases and their boundaries (see \cite{supp} and \cite{Zhuang2015} for similar phase diagrams obtained using entanglement entropy). (b) $\epsilon = 0.075$, a large value that causes both the samples in the PM and IC phases to be grouped into one cluster, while in the FM phase three clusters are always identified.}
	\label{fig:Clock_diffmaps}
\end{figure}

The incommensurate phase and its boundary can be numerically identified using entanglement entropy \cite{calabrese_entanglement_2004,Zhuang2015}, a quantity hard to measure in large experimental systems \cite{Islam_2015}. Here we show that using just the measurement samples of the spin operators $\{\sigma_j\}$, the diffusion map is able to produce a phase diagram in an unsupervised manner that well matches the one obtained using entanglement entropy. Using an intermediate value of $\epsilon$, the number of clusters identified by the diffusion map identifies all three phases of the $\mathbb{Z}_3$ TFIM, as shown in  Fig.\,\ref{fig:Clock_diffmaps}(a). This result can be understood as follows: (1) Deep in the PM phase the measurement samples are approximately drawn from a uniform probability distribution of every possible spin configuration. As a result, two different samples will have on average $N/2$ spins in different configurations. Since in practice the number of samples is often much smaller than the number of spin configurations, each sample will be treated as a separate cluster if $\epsilon \ll 1/\ln\tau $. The number of clusters will thus be close to the number of samples. (2) Deep in the FM phase, the ground state undergoes spontaneous symmetry breaking, resulting in one of the 3 FM states, each being ordered in a different direction. The samples drawn from each of the 3 FM ordered states should have small distances between each other while the samples drawn from different FM ordered states have very large distances. With $1/(N\ln\tau) \ll \epsilon \ll 1/\ln\tau$, the number of clusters identified will be close to 3. (3) The samples drawn from the IC phase should have a rather diverse set of distances and the number of clusters identified by the diffusion map should vary strongly depending on the parameters of the Hamiltonian. Note that no fine tuning of $\epsilon$ is needed to learn the incommensurate phase. In addition, we have shown in the supplementary material \cite{supp} that one can identify all three phases equally well if the samples are drawn from measurements of the $\tau$ operators instead.

The diffusion map also allows unsupervised learning of the order of the discrete symmetry group underlying a symmetry breaking phase transition. For a large enough $\epsilon$ ($\sim 1/\ln\tau$), samples in the PM and IC phases will be identified as a single cluster while samples in the FM phase are cleanly sorted into 3 clusters as a result of the $Z_3$ spontaneous symmetry breaking [see Fig.\,\ref{fig:Clock_diffmaps}(b)]. Contrast this to the k-means clustering algorithm used frequently in unsupervised machine learning \cite{Lloyd_1982}, where the number of clusters has to be guessed or predicted using a priori knowledge of the data. 

\sect{Learning valence-bond solid phase transitions} Valence-bond solids (VBS) are important in condensed matter physics and quantum information as they are closely related to quantum spin liquids \cite{balents_spin_2010}, symmetry protected topological order \cite{Affleck_1987,Chen_2013}, tensor network states \cite{Or_s_2014}, and cluster states for quantum computing \cite{Verstraete_2004}. Because VBS cannot be identified using an order parameter linear in spin operators, this is another scenario where unsupervised learning methods such as PCA will fail while diffusion maps are useful. As a specific example, we consider a spin-1/2 chain with nearest-neighbor and next-nearest-neighbor antiferromagnetic Heisenberg interactions \cite{Majumdar_1969}, commonly known as the $J_1$-$J_2$ model, with the Hamiltonian $H_2=\sum_{j=1}^{N} (J_1 \bm{S}_j \cdot \bm{S}_{j+1} +J_2 \bm{S}_j \cdot \bm{S}_{j+2}$). For simplicity we set $J_1=1$ below. This is a paradigmatic model exhibiting VBS order, where the ground state at $J_2=0.5$ is exactly solvable and made of two degenerate VBS (dimer) states, corresponding to two different ways of pairing neighboring spins into spin-1/2 singlets (one with $\bm{S}_i+\bm{S}_{i+1}=0$ for odd $i$ and the other for even $i$). At around $J_2\approx 0.3$ \cite{J1J2_tonegawa1987ground}, the system is expected to undergo a phase transition from the Luttinger liquid to the VBS phase.

\begin{figure}[htp]
	\includegraphics[width=\columnwidth]{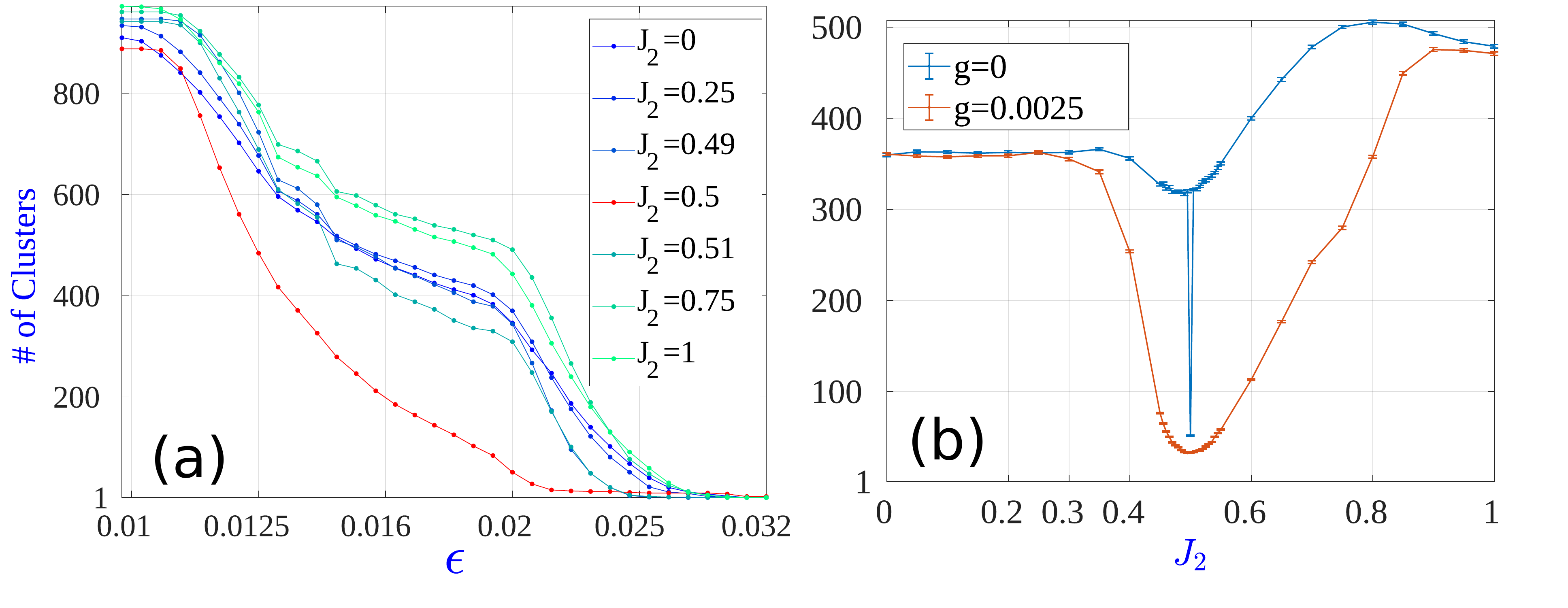}
	\caption{(a) The number of clusters found by diffusion maps when applied to the measurement samples of the $J_1$-$J_2$ model (see $H_2$) as a function of $\epsilon$. 1000 samples are used for each value of $J_2$ ($J_1=1$), obtained from exact diagonalization of $H_2$ with $N=24$ spins. (b) An intermediate $\epsilon$ ($\epsilon=0.02$) is chosen to show that the number of clusters drops at the onset of the valence-bond phase transition with or without a symmetry breaking perturbation. The error bar shows the standard error of the mean from $90$ repeated sampling processes.}
	\label{J1J2_Ndegen}
\end{figure}

As expected, we find no signatures of the VBS phase transition and no special behavior at $J_2=0.5$ using PCA and k-means clustering \cite{supp}. With diffusion maps, as shown in Fig.\,\ref{J1J2_Ndegen}(a), we can easily identify the exactly solvable point of $J_2=0.5$, where the ground state has spontaneous translational symmetry breaking even for a finite system size. To better see the phase transition at $J_2\approx 0.3$, we add a small symmetry breaking perturbation of the form  $H^{\prime}=g \sum^N_{j=1} [1-(-1)^j] \bm{S}_j \bm{S}_{j+1}$ that breaks the $Z_2$ translational symmetry of $H_2$ for the small system size our calculation is limited to. As shown in Fig.\,\ref{J1J2_Ndegen}(b), we see a large dip in the number of clusters starting around $J_2\approx0.3$ for an intermediate $\epsilon$ value. This is because when the VBS phase transition occurs, symmetry breaking in the ground states picks one of the dimer configurations, while away from the phase transition point, both dimer configurations coexist. The probabilities of finding two samples with a small distance from each other is much lower if the two samples are from different dimer configurations than from a single dimer configuration. Thus the number of clusters identified by the diffusion map drops when translational symmetry breaking takes place. This argument can be made mathematically precise for the exact degeneracy point of $J_2=0.5$, where finite-size effects are irrelevant \cite{supp}. We expect a similar picture for general VBS phase transitions.

\sect{Learning many-body localization} A 1D quantum system with tunable disorder can exhibit a dynamical phase transition from a thermal, ergodic phase to a many-body localized (MBL) phase \cite{nandkishore2015many}. There is no simple order parameter to detect the MBL phase transition. Theoretically, one can use the inverse participation ratio, level statistics, or entanglement entropy to detect a MBL phase transition \cite{MBL_Pal_Huse}. However, these quantities are difficult to obtain experimentally. A more practical way to detect MBL is to use quench dynamics. For example, one can measure local observables after a long-time evolution from some initial, simple-to-prepare product state. Here we show that the diffusion map method can learn the thermal-to-MBL phase transition using the measurement samples obtained in quench dynamics experiments unsupervised. This is different from existing machine learning studies of MBL that require supervised learning \cite{Schindler_2017,Hsu2018}.

\begin{figure}
	\includegraphics[width=\columnwidth]{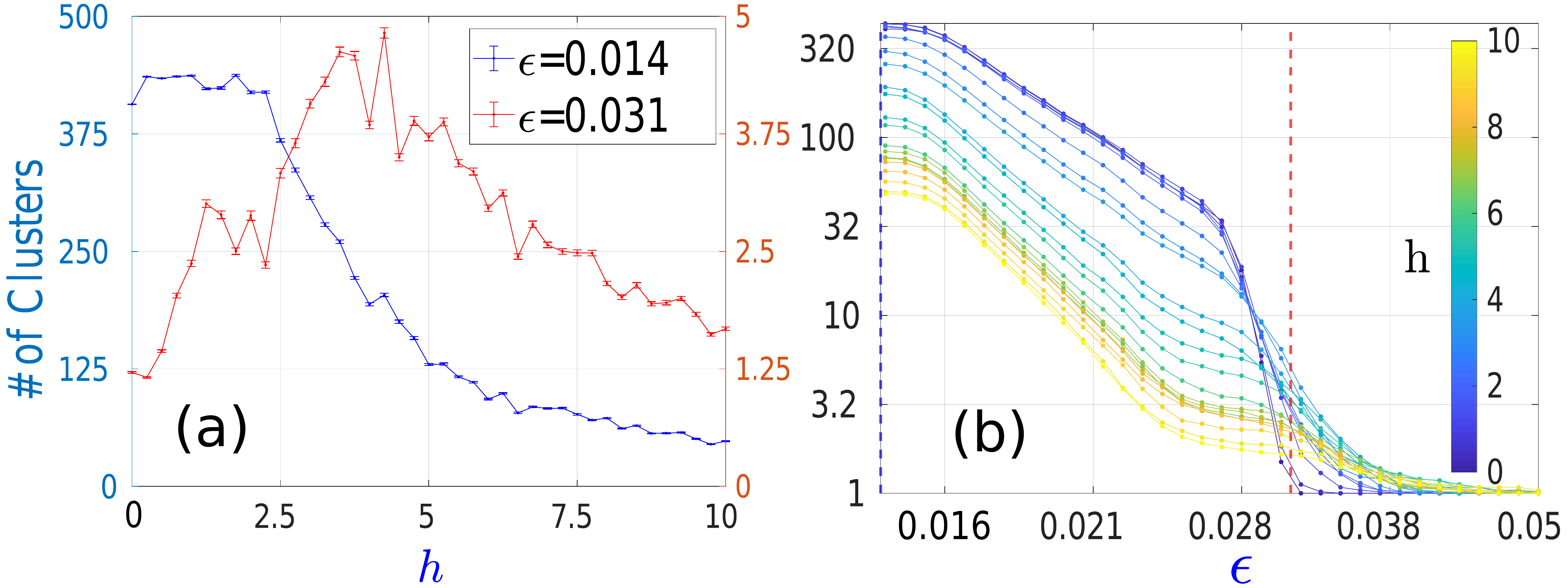}
	\caption{Number of clusters learned by diffusion maps on the measurement samples of the long-time dynamical state of $H_3$, averaged over 50 disorder realizations. 500 samples are obtained for $N=16$ spins using exact diagonalization. (a) A small $\epsilon$ value (blue curve) leads to the number of samples decreasing rapidly with increased disorder, while the peak of the number of clusters with an intermediate $\epsilon$ value (red curve) reveals the approximate location of the thermal-to-MBL phase transition. The error bars show the standard error of the mean calculated over $50$ different disorder realizations and $50$ repeated sampling processes. (b) There exists a range of intermediate $\epsilon$ values where the number of clusters peaks near the critical disorder strength.}
	\label{MBL_fig}
\end{figure}

As an example, we study a paradigmatic model exhibiting the thermal-MBL phase transition, i.e. the spin-1/2 Heisenberg model with a random field \cite{MBL_Pal_Huse,Schindler_2017}, with the Hamiltonian $H_3= \sum^N_{i=1}{ J \bm{S}_i \cdot \bm{S}_{i+1} +h_i S_i^z}$. Here  $h_i \in (-h, h)$ is a random number drawn from a uniform distribution and $h$ denotes the disorder strength. It has been found numerically that the thermal-MBL phase transition takes place at the critical disorder strength $h_c \approxeq 3.5 \pm 1.0$ with $J=1$. We perform quench dynamics using an initial state with $\langle S_i^z\rangle =\frac{(-1)^i}{2}$ and measure all $S_i^z$ ($i=1,2,\cdots,N$) after a long time ($t= 10^4/J$). As mentioned before, since the number of unique samples decreases with increasing disorder strength, we see that the diffusion map with a small $\epsilon$ can already indicate the onset of MBL [Fig.\,\ref{MBL_fig}(a)]. But we can learn more about where the thermal-MBL phase transition occurs by using an intermediate value of $\epsilon$. As shown in Fig.\,\ref{MBL_fig}, the number of clusters identified by the diffusion map is actually peaked near $h_c$ for a range of intermediate values of $\epsilon$. Intuitively, this is because deep in the thermal phase, the samples are scattered across configuration space with similar distances between each other. They are unlikely to form many small clusters. On the other hand, deep in the localized phase the samples are already clustered around the initial state, and the number of clusters should also be small. However, near thermal-MBL phase transitions, the samples form a number of scattered, small clusters in configuration space due to the competition of disorder and ergodicity. As a result, we expect to see the most clusters near the phase transition with intermediate $\epsilon$. We have also observed this behavior for other spin models exhibiting MBL \cite{supp}, suggesting that diffusion maps are widely applicable in learning MBL. Note that for learning MBL, we need to use a different kernel in the diffusion map to accommodate for the large disparity in the densities of measurement samples in configuration space between the thermal and MBL phases. More precisely, we use $K_{ij}/(\sum_k K_{ik} \sum_k K_{kj})$ in place of $K_{ij}$ \cite{supp}. This kernel normalization is a standard practice when applying diffusion maps to data with large variations in density \cite{Coifman2006}. We emphasize that this modified kernel is not fine tuned and can learn the aforementioned incommensurate and VBS phase transitions equally well.

\sect{Conclusion and Outlook} We have shown that the diffusion map is a general and versatile method for unsupervised learning of various complex quantum phase transitions. Compared to traditional data analysis methods, diffusion maps use the full statistics of the measurement samples which contains information of spin correlations at all orders, and thus offers more knowledge of the measured quantum states without demanding more data. Because a quantum phase transition is always accompanied by abrupt changes in correlations of local observables at certain orders, if an appropriate local measurement basis is used (which can often be guessed using knowledge of the experimental Hamiltonian), the diffusion map method has a high chance of revealing the underlying quantum phase transition from the measurement data. The limitations of diffusion maps in learning quantum phases, however, are far from clear. For example, here we mainly focus on the number of clusters identified by the fictitious diffusion process, but the diffusion map often involves a mapping of the samples to the low-dimensional subspace spanned by the leading eigenvectors of the $P$ matrix. Such a mapping could reveal more information regarding the geometry or topology of the samples in configuration space, which may be particularly useful in learning certain topological phases  \cite{Rodriguez-Nieva2019} or symmetry-protected topological phases. In addition, typical diffusion maps use Euclidean distance and Gaussian kernels, but some exotic phases of matter may require the use of very different distance metrics or kernel functions to be learned. Furthermore, can we learn more with diffusion maps if we use samples obtained by measurements in more than one local basisThis may be particularly useful if the studied system contains multiple phases of distinct nature, or for distinguishing classical and quantum phase transitions. And finally, how will other nonlinear dimensionality reduction methods widely used in machine learning, such as t-SNE \cite{t-SNE} and DBSCAN \cite{DBSCAN} compare to the performance of diffusion maps when applied to quantum phase detection?

\begin{acknowledgments}
We thank Lincoln Carr, Eliot Kapit, Cecilia Diniz Behn, and Bo Wu for enlightening discussions related to this work, and the HPC center at Colorado School of Mines for providing computational resources needed in carrying out this work. AL and ZXG acknowledge funding support from the NSF RAISE-TAQS program under Grant No. CCF-1839232.
\end{acknowledgments}

\bibliographystyle{apsrev4-2}
\bibliography{refs}

\end{document}

\title{Supplementary Material for ``Unsupervised machine learning of quantum phase transitions using diffusion maps''}

\author{Alexander Lidiak}
\email{alidiak@mines.edu}
\affiliation{Department of Physics, Colorado School of Mines, Golden, Colorado 80401, USA}
\author{Zhexuan Gong}
\email{gong@mines.edu}
\affiliation{Department of Physics, Colorado School of Mines, Golden, Colorado 80401, USA}
\affiliation{National Institute of Standard and Technology, Boulder, Colorado 80305, USA}

\date{\today}

\maketitle

In this supplementary material, we will show additional supporting results for how diffusion maps can learn incommensurate phases (section I), valence-bond solid phases (section II), and many-body localization (section III).

\section{Learning incommensurate phases}

In this section, we first show that principle component analysis (PCA) together with k-means clustering is unable to learn the incommensurate phase of the $\mathbb{Z}_3$ chiral transverse-field Ising model (see $H_1$ in the main text) and its  boundary. We perform PCA on the same collection of measurement samples used for the diffusion map in Fig\,2 of the main text and extract the projection of the sample set onto the first two principle components. We then apply a k-means clustering algorithm to associate each sample $\bm{X}_i$ with an index $L_i=1,2,\cdots,k$. In an attempt to identify all three phases within the samples, we manually set the number of clusters to $k=3$ (note that the diffusion map does not require such a priori knowledge of the number of distinct phases in the data). We then average $L_i$ for samples belonging to a particular set of Hamiltonian parameters ($f$, $\theta$), and obtain a phase diagram using this averaged index [see Fig.\,\ref{CCM_Kmeans}(a)]. While we can identify the ferromagnetic phase and its boundary, there is no clear identification of the paramagnetic or incommensurate phases.

We have also used an auto-encoder included in MATLAB to perform nonlinear dimensionality reduction of the measurement data in substitution of PCA. The auto-encoder trains an artificial neural network to retain as much information of the sample data as possible with two latent variables onto which we then encode each measurement sample (similar to projecting onto the two principle components obtained via PCA). Applying the same k-means clustering algorithm with $k=3$ leads to a phase diagram shown in Fig.\,\ref{CCM_Kmeans}(b), which again is unable to identify the incommensurate phase. This is because the incommensurate phase cannot be identified using a single linear or nonlinear function of the measured observables, as the spin configurations in this phase vary strongly with the system parameters ($\theta$ and $f$). The diffusion map method, however, does not try to reduce the dimensionality of the data directly. Instead, it detects the change in the distribution of the measurement samples in configuration space which is often linked to a phase transition.

\begin{figure} [htp]
	\includegraphics[width=0.49\textwidth]{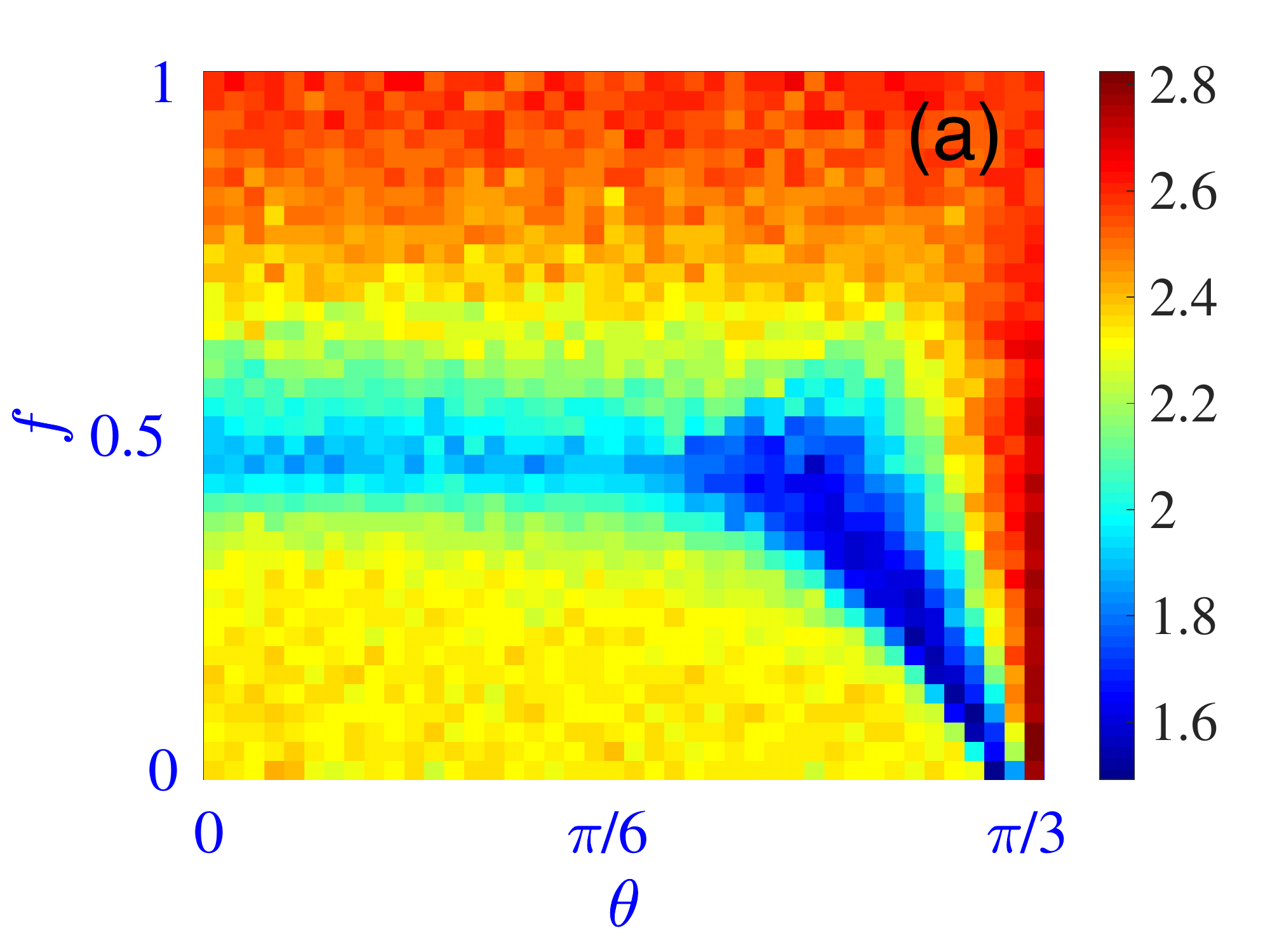} 
	\includegraphics[width=0.49\textwidth]{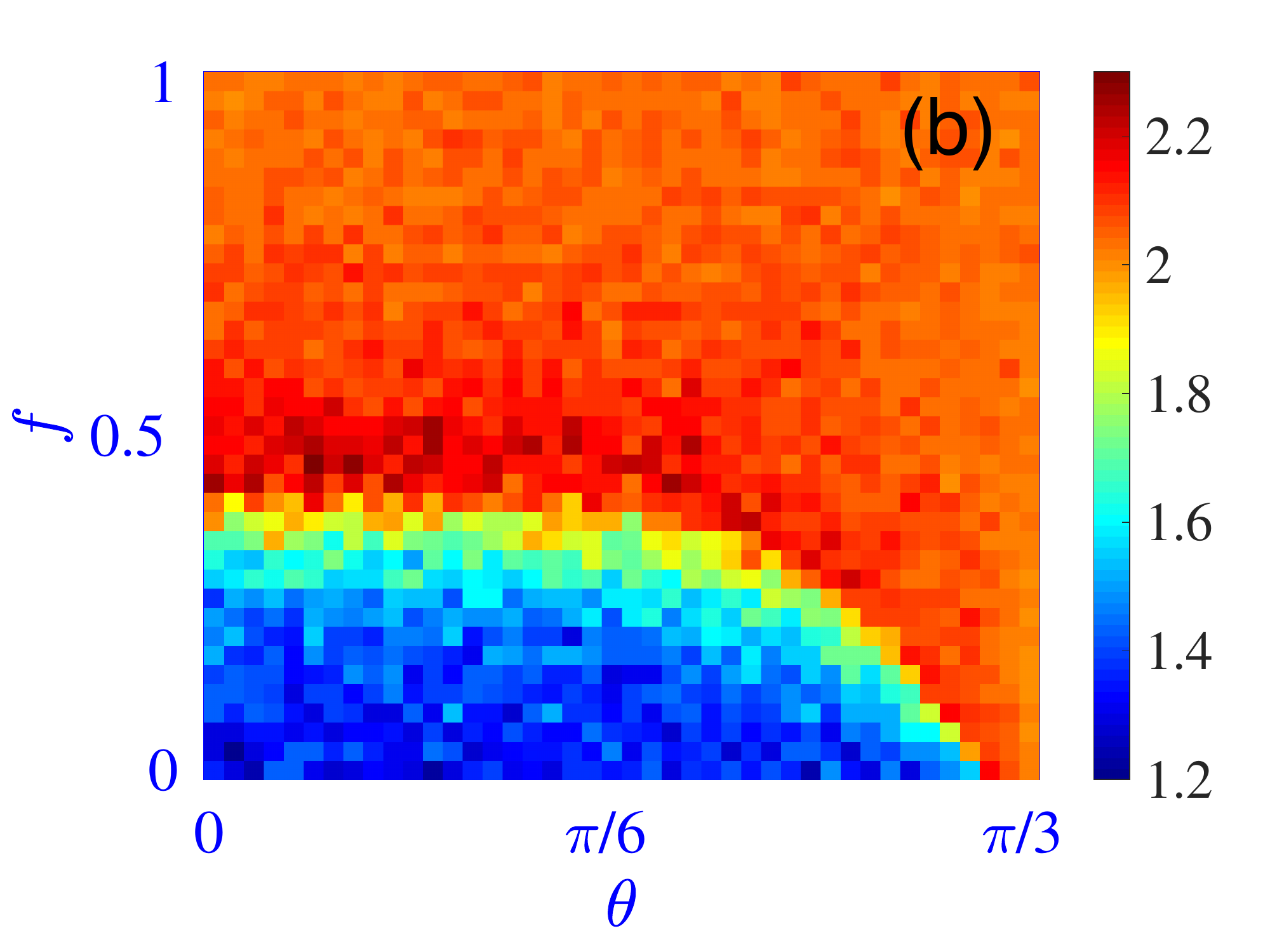}
	\caption{Phase diagrams generated by PCA (a) and auto-encoder (b) applied on the same measurement samples used in Fig.\,2 of the main text. The color represents the index attached by a $k=3$ k-means clustering algorithm on the projected/compressed measurement sample data, averaged over all measurement samples of the ground state of a particular Hamiltonian.}
	\label{CCM_Kmeans}
\end{figure}

In addition, we show the phase diagram obtained using diffusion maps for a much larger system size than that used in Fig.\,2 of the main text. We use the OpenMPS library \cite{Jaschke_2018} to variationally find the ground state of $H_1$ (see main text) for $N=100$ spins using a bond dimension of 200. We then use the Metropolis-Hastings Monte Carlo sampling method \cite{Intro_MC_Methods} to efficiently generate measurement samples of $\sigma_j$ ($j=1,2,\cdots,N$) using the matrix product state (MPS) ansatz that approximates the ground state. We then perform diffusion maps with an intermediate value of $\epsilon$. As shown in Fig.\,\ref{fig:Clock_diffmaps}(a), the phase diagram obtained using the number of clusters identified by the diffusion map matches well with Fig.\,2(a) in the main text and that obtained using the half-system entanglement entropy shown in Fig.\,\ref{fig:Clock_diffmaps}(b). Note that the noises in the ferromagnetic phases in Fig.\,\ref{fig:Clock_diffmaps}(b) are due to very small energy gaps in the ground state manifold for $N=100$ spins such that the variational MPS algorithm may also be subject to spontaneous symmetry breaking.

\begin{figure} [h!]
    \includegraphics[width=0.49\textwidth]{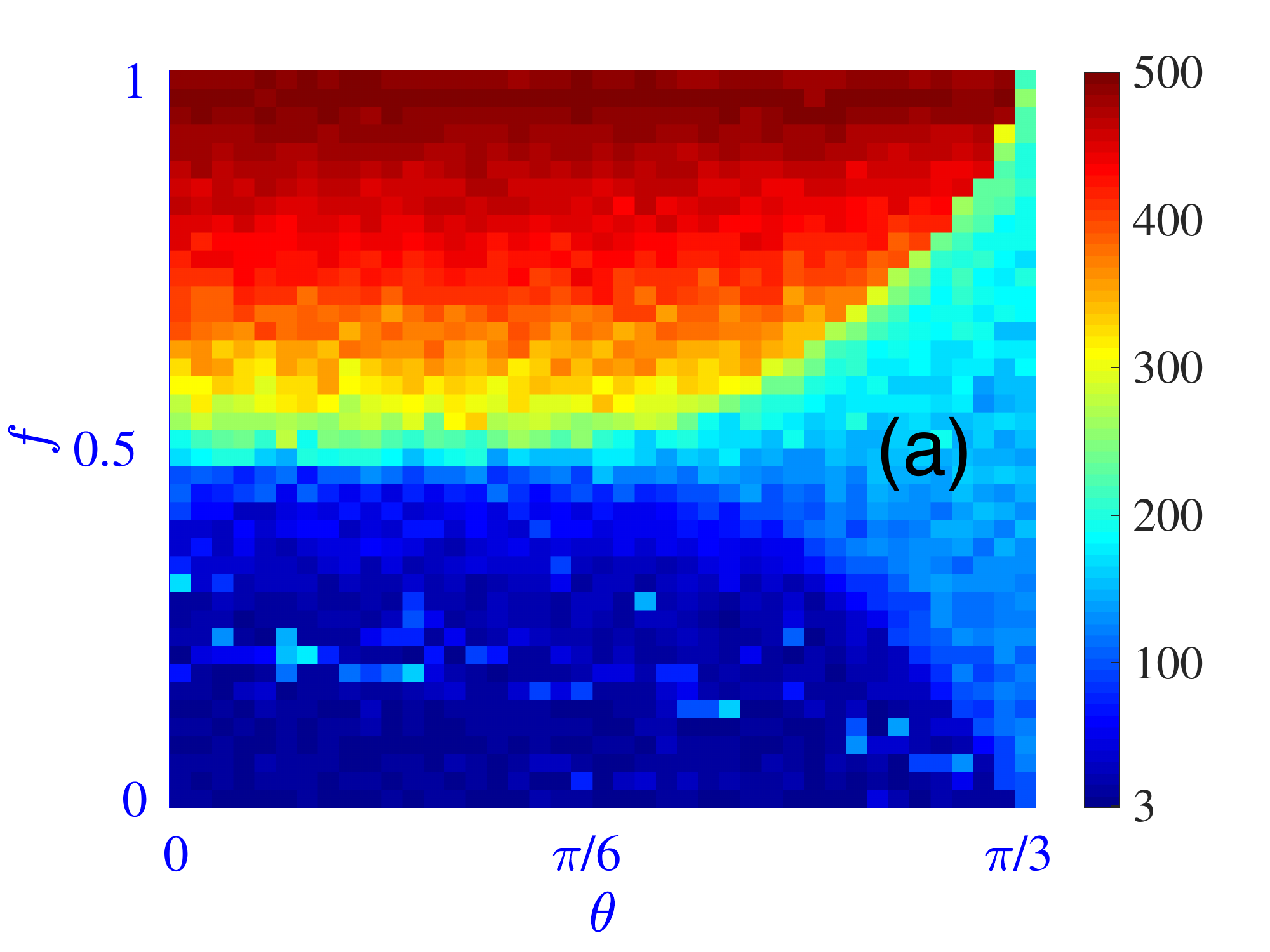}  \includegraphics[width=0.49\textwidth]{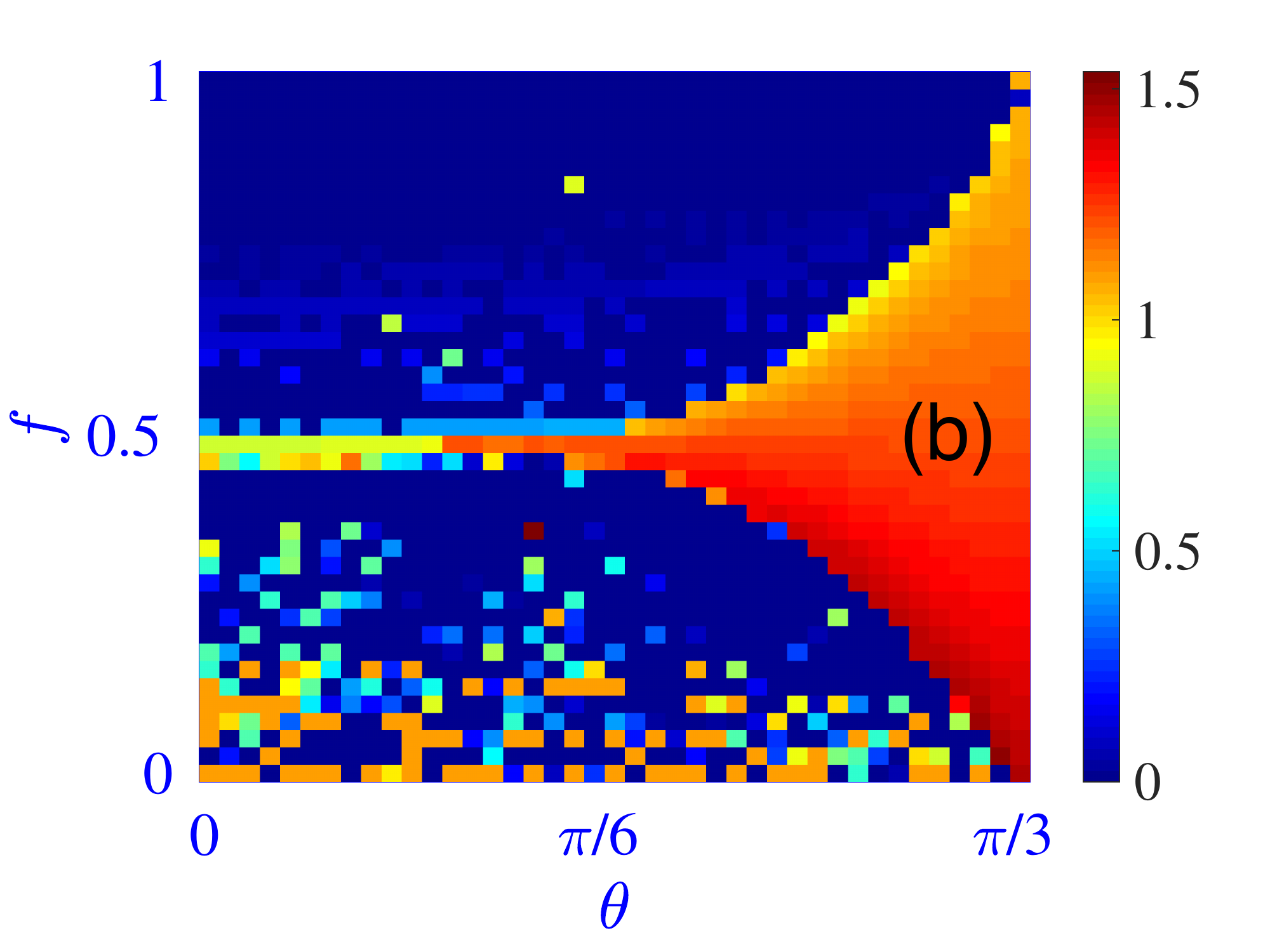}

	\caption{(a) Ground-state phase diagram of $H_1$ (see main text) obtained by (a) performing diffusion maps with $\epsilon=0.004$, $\delta=10^{-2.5}$ on 500 measurement samples generated for each Hamiltonian using Monte Carlo sampling. (b) calculating the half-system entanglement entropy using MPS methods.}
	\label{fig:Clock_diffmaps}
\end{figure}

\section{Learning valence-bond solid phases}
In this section, we will first show that for the $J_1$-$J_2$ model discussed in the main text ($H_2$), PCA and k-means clustering cannot detect the formation of valence-bond solids (VBS) or the spontaneous symmetry breaking at $J_2=0.5$. We first perform PCA on the same measurement samples used in Fig.\,3 of the main text. As shown in Fig.\,\ref{J1J2_kmeans}(a), the efficacy of dimensionality reduction is poor in this case, with many principle components contributing significantly to the variance of the data. Moreover, the first principle component only identifies an anti-ferromagnetic order, which is irrelevant for the VBS phase transition. Keeping the first two principle components, we apply a k-means clustering algorithm with $k=2$ and plot the average index as a function of $J_2$ [Fig.\,\ref{J1J2_kmeans}(b)]. There is no clear signature of a phase transition happening at $J_2\approx 0.3$ and no indication of the spontaneous translational symmetry breaking at $J_2=0.5$. We have also used auto-encoder in place of PCA and find it performs no better.

\begin{figure}[h]
		\includegraphics[width=0.48\textwidth]{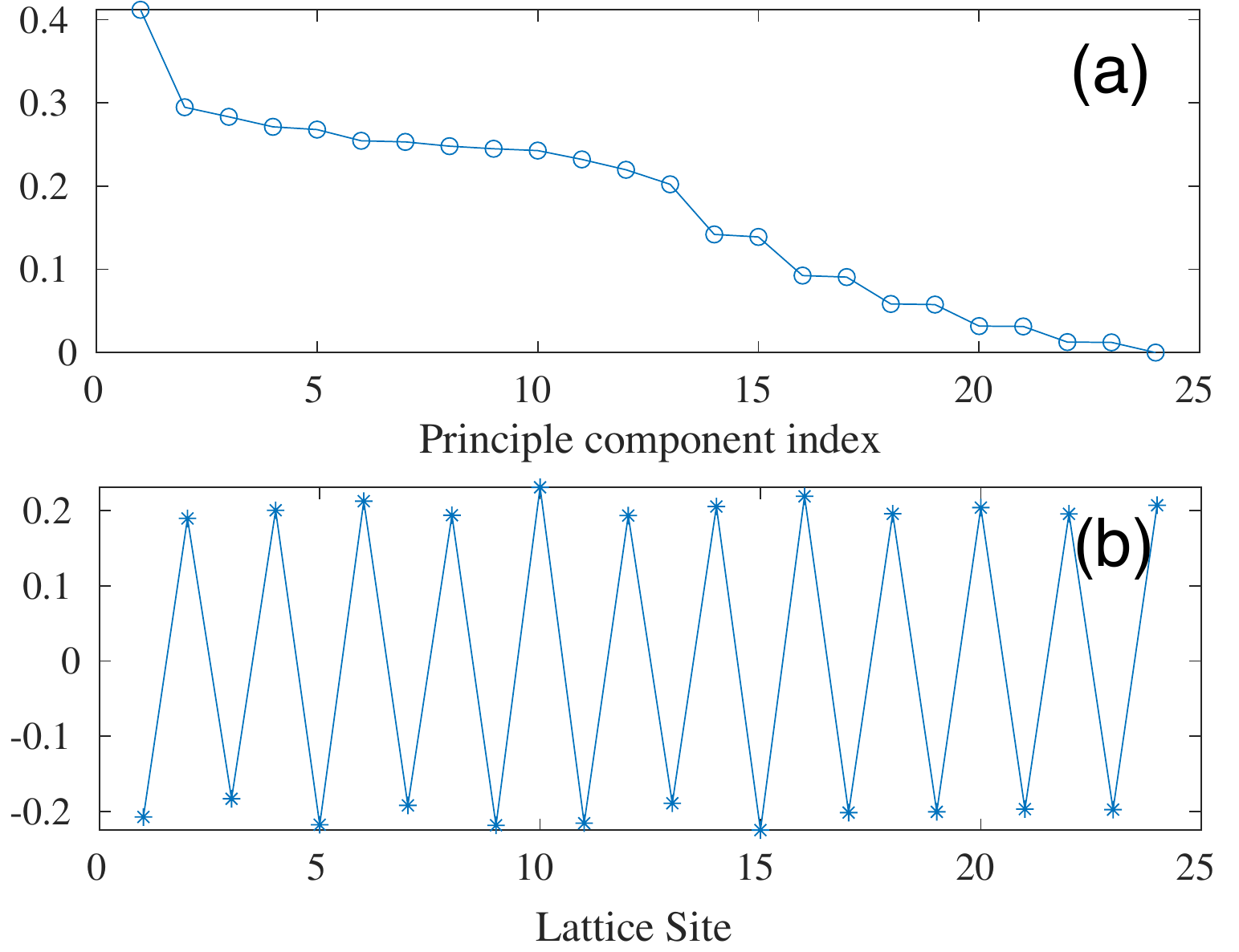}
		\includegraphics[width=0.5\textwidth]{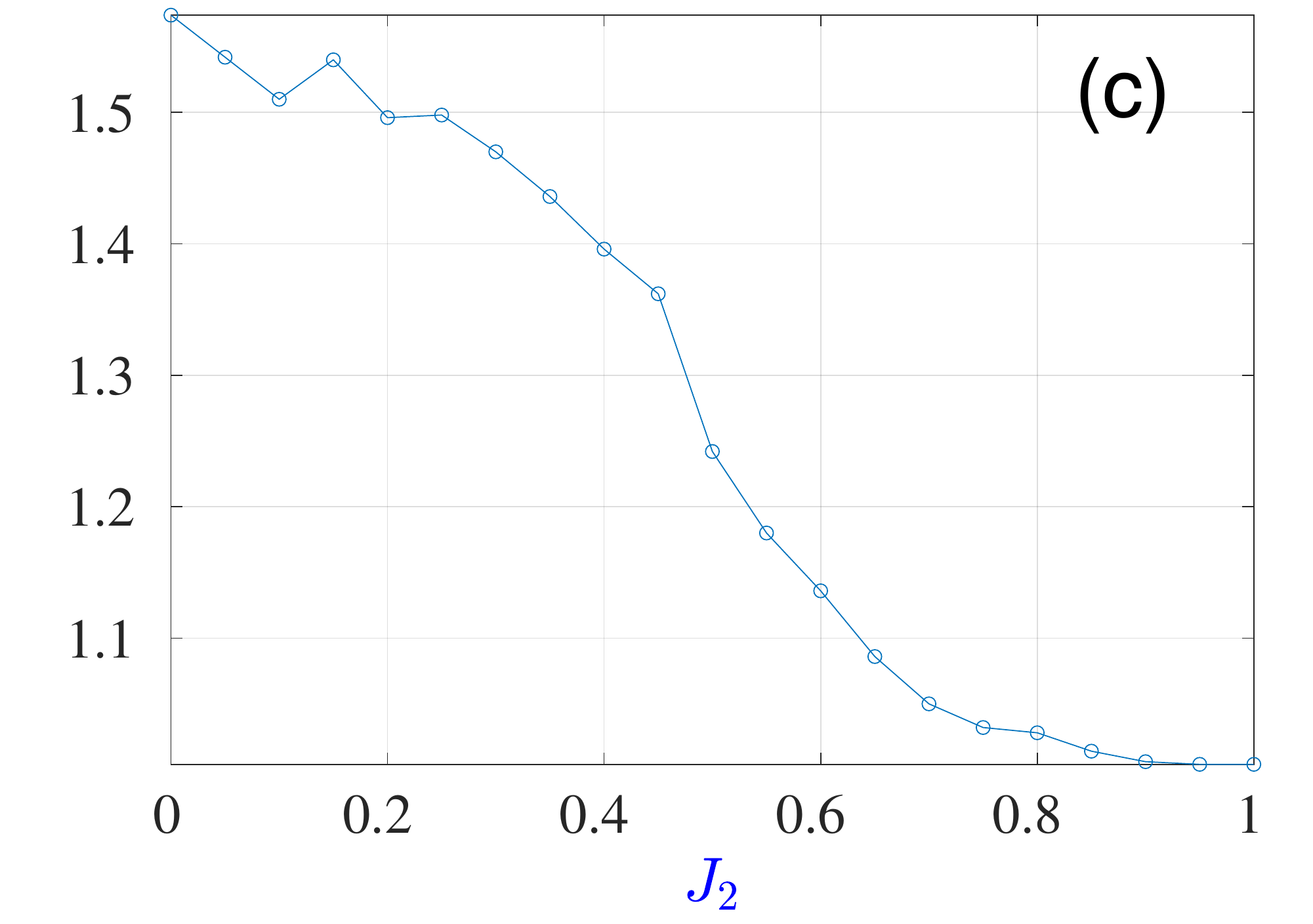}
	\caption{ (a) Principle values of PCA applied to the same measurement data used in Fig.\,3 of the main text. (b) Coefficients of the first principle component as a linear combination of each spin's magnetization, which indicates that the first principle component is an anti-ferromagnetic order parameter. (c) The average index assigned by k-means clustering with $k=2$ to the measurement samples projected onto the first two principle components, as a function of $J_2$.}
	\label{J1J2_kmeans}
\end{figure}

Next, we will explain in details the sharp drop in the number of clusters at $J_2=0.5$ identified by the diffusion map with an intermediate value of $\epsilon$. First, we point out that at $J_2=0.5$, we only obtain one of the two degenerate dimer states (one with $\bm{S}_i+\bm{S}_{i+1}=0$ for all odd $i$s and the other for all even $i$s) as the ground state numerically, which is also expected in an actual experiment due to spontaneous symmetry breaking. A small deviation from $J_2=0.5$ will lift the degeneracy of the two dimer states and the ground state becomes approximately a superposition of the two dimer states, which we will call the `combined dimer state' below. Thus to see why the number of clusters identified by diffusion maps suddenly drops at $J_2=0.5$, we can compare the results of diffusion maps on a single dimer state with that of the combined dimer state. This comparison is shown in Fig.\,\ref{J1J2_Ndegen}, which is very similar to how the $J_2=0.5$ curve compares to the $J_2$ close to $0.5$ curves in Fig.\,3(a) of the main text.

To understand why the number of clusters for the single dimer state decreases much more rapidly than the combined dimer state, let us start from the following analysis: If we get two random measurement samples from the combined dimer state, then there is $1/2$ probability that both samples are drawn from either one of the single dimer states (Case I), and $1/2$ probability that the two samples are drawn from two different single dimer states (Case II). The intuition is that the probability of finding two samples that are close to each other is much smaller in Case II than in Case I. For example, in Case I, for a given first sample, the chance of getting the second sample that has zero distance (i.e. identical) to the first sample is always $1/2^{N/2}$. But in Case II, the chance of getting an identical sample is $2/2^N$ because there are only two samples that can be obtained from both of the dimer states (which are the two perfect antiferromagnetic states), and each only appears with a probability of $1/2^{N/2}$. As a result, we can largely ignore case II in finding two samples close to each other. As we will show below, in the thermodynamic limit ($N\rightarrow\infty$), case II can be completely ignored except when we are considering two samples with exactly half of the spins in different directions. After ignoring case II, the probability of finding two samples with $k$ spins different for the combined dimer state, denoted by $P_c(k)$, is only half that of the single dimer state, denoted by $P_s(k)$, in the large $N$ limit.

The above analysis can be made precise mathematically. We find that $P_s(k)=\binom{\frac{N}{2}}{\frac{k}{2}}/2^{\frac{N}{2}}$ and $P_c(k)=[2^{\frac{N}{2}-1} \binom{\frac{N}{2}}{\frac{k}{2}}+\binom{N}{k}]/2^{N}$. We have plotted the ratio $P_s(k)/P_c(k)$ in Fig.\,\ref{J1J2_Ndegen}(b) for $N=24$ and $N=1000$. In the $N\rightarrow\infty$ limit, one can show analytically that $P_s(k)/P_c(k)=2$ for $k\ne N/2$ and $P_s(k)/P_c(k)=2/(1+\sqrt{2})\approx 0.83$ for $k=N/2$. Note that both $P_s(k)$ and $P_c(k)$ are symmetric around $k=N/2$.

Because the probability of finding two samples for most distances in the single dimer state is two (or close to two for finite $N$) times larger than that in the combined dimer state, the number of clusters for the single dimer state will decrease with the cluster radius (proportional to $\epsilon$) at twice the rate of that for the combined dimer state. This twice as fast decay is what we observe in Fig.\,\ref{J1J2_Ndegen}(a) as well as Fig.\,3 of the main text which exhibits similar physics.

\begin{figure}

        \includegraphics[width=0.49\textwidth]{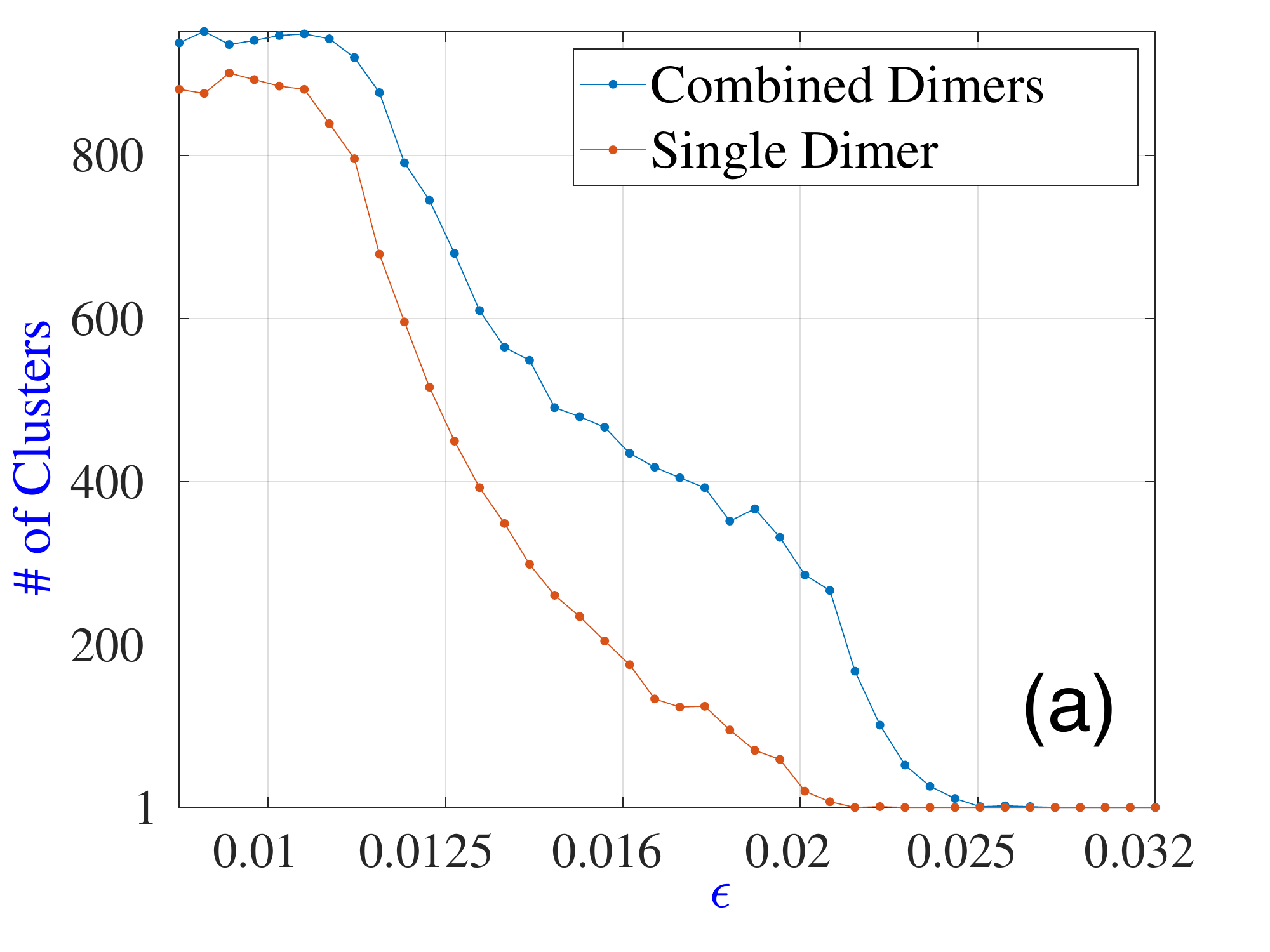}\hfill
	    \includegraphics[width=0.5\textwidth]{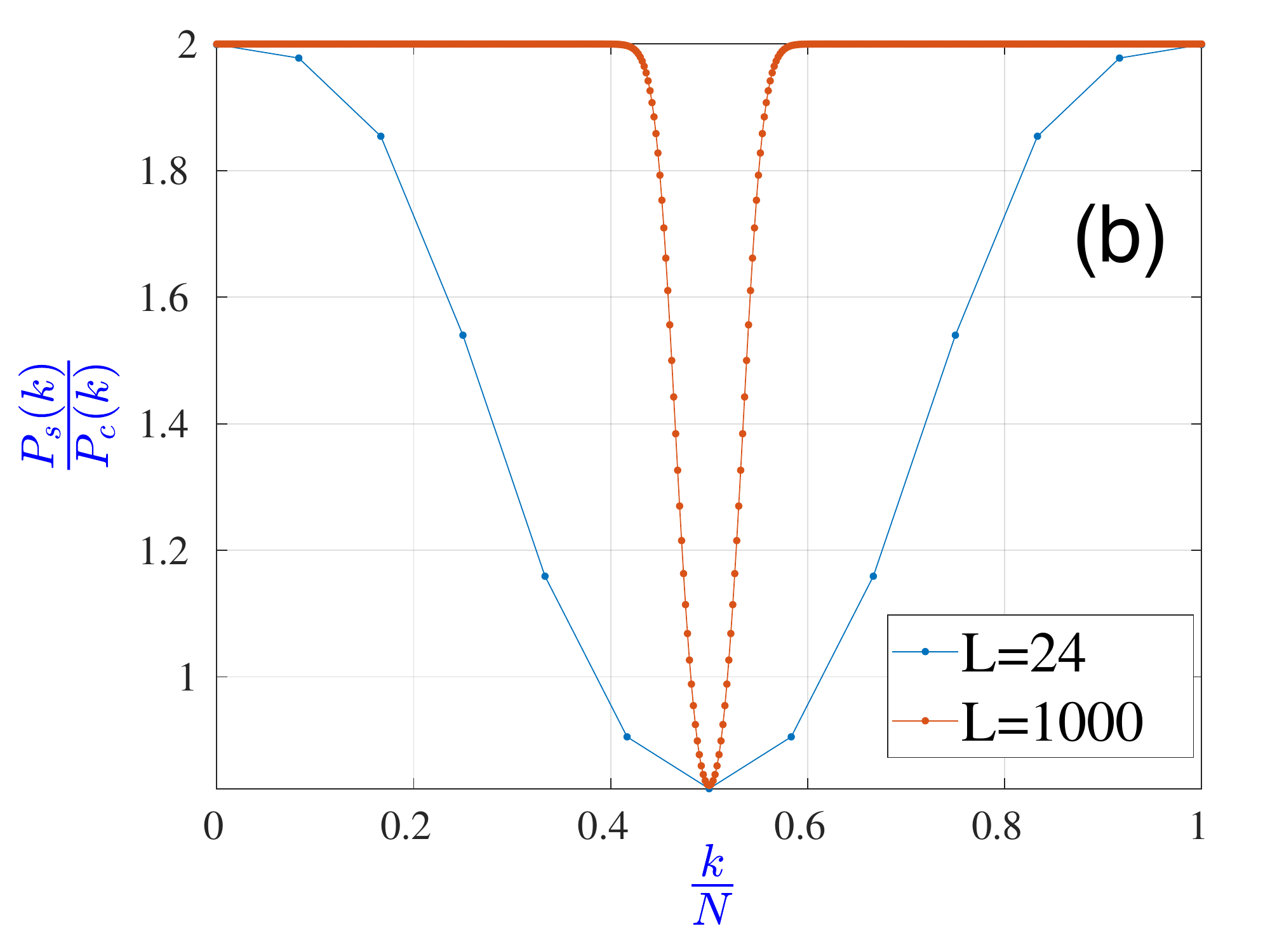}
	\caption{(a) The number of clusters identified by the diffusion map as a function of $\epsilon$ used on the samples drawn from the single dimer state versus the combined dimer state with $N=24$. (b) The ratio $P_s(k)/P_c(k)$ of the probability of finding two samples with $k$ spins different in the single dimer state to that in the combined dimer state.}
	\label{J1J2_Ndegen}
\end{figure}

\section{Learning many-body localization}

We will first explain why we use a different kernel in the diffusion map for learning many-body localization. In the many-body localized phase, the measurement samples will be clustered around the point of the initial state, while in the thermal phase, the measurement samples will be scattered across the configuration space. This leads to a large variation in the density of samples in the configuration space, and the original Gaussian kernel (see main text) results in the number of clusters always decaying with increasing disorder strength for all values of $\epsilon$. This obscures the physical picture in that when the system goes from the thermal to the MBL phase, samples of the long-time dynamical states will start to form many small clusters due to the interplay of ergodicity and localization. To address this issue, we adopt a different kernel in the diffusion map that is widely used in data science \cite{Coifman2006}, corresponding to an extra normalization of the Gaussian kernel (see the $\alpha=1$ case in Refs.\,\cite{Coifman2006,Rodriguez-Nieva2019}), i.e. we will use $K^{\prime}_{ij}=K_{ij}/(\sum_k K_{ik} \sum_k K_{kj})$ as the kernel in the place of $K_{ij}$. This extra normalization performed on $K_{ij}$ eliminates the density dependence of the samples in extracting the structure of the samples in configuration space \cite{Coifman2006}. In the context of the thermal-to-MBL phase transition, we see that if the samples are drawn from a disordered state, then most of the samples are close to each other, with only a small amount of samples far away from the rest. Using the kernel given by $K^{\prime}_{ij}$, the transition probabilities between samples that are closely packed (far apart) will be reduced (increased), effectively evening the density of samples in the configuration space. We note that for samples without a big variation of density (e.g. those generated from the $J_1$-$J_2$ and chiral clock models we studied), using either $K_{ij}$ or $K^{\prime}_{ij}$ makes little difference.

Finally, we show how diffusion maps can learn the many-body localization phase transition of a different model than the one studied in the main text. This model is a disordered spin-1/2 transverse field Ising chain with next-nearest neighbor interaction, with Hamiltonian:
\begin{equation}
    H= - \sum^{N-1}_{i=1} J_i \sigma_i^z \sigma^z_{i+1} +J_2 \sum^{N-2}_{i=1} \sigma_i^z \sigma^z_{i+2}+h \sum^N_{i=1} \sigma^x_i. \label{H}
\end{equation}
Here the nearest neighbor Ising couplings are disordered as $J_i=J+\delta J_i$ with $\delta J_i$ drawn from a uniform random distribution $[-\delta J,\delta J]$. According to Ref.\,\cite{J1rand_MBL}, this Hamiltonian undergoes a thermal to MBL phase transition near $\delta J_c=3.81 \pm 0.04$ (depending also on the energy density of the initial state) when $J=1$ and $\frac{h}{2}=J_2=0.3$. In Fig.\,\ref{J1_rand_fig}(a), we show the number of clusters of the diffusion map applied to the measurement samples of $\{\sigma_i^z\}$ drawn from the long-time dynamical state of an initial antiferromagnetic spin state as a function of $\epsilon$. We find that for intermediate values of $\epsilon$, the number of clusters shows a peak around $\delta J_c$ [Fig.\,\ref{J1_rand_fig}(b)]. This provides further evidence that diffusion maps are able to provide signatures of thermal to MBL phase transitions in general.

\begin{figure}[h]
	\includegraphics[width=0.49\linewidth]{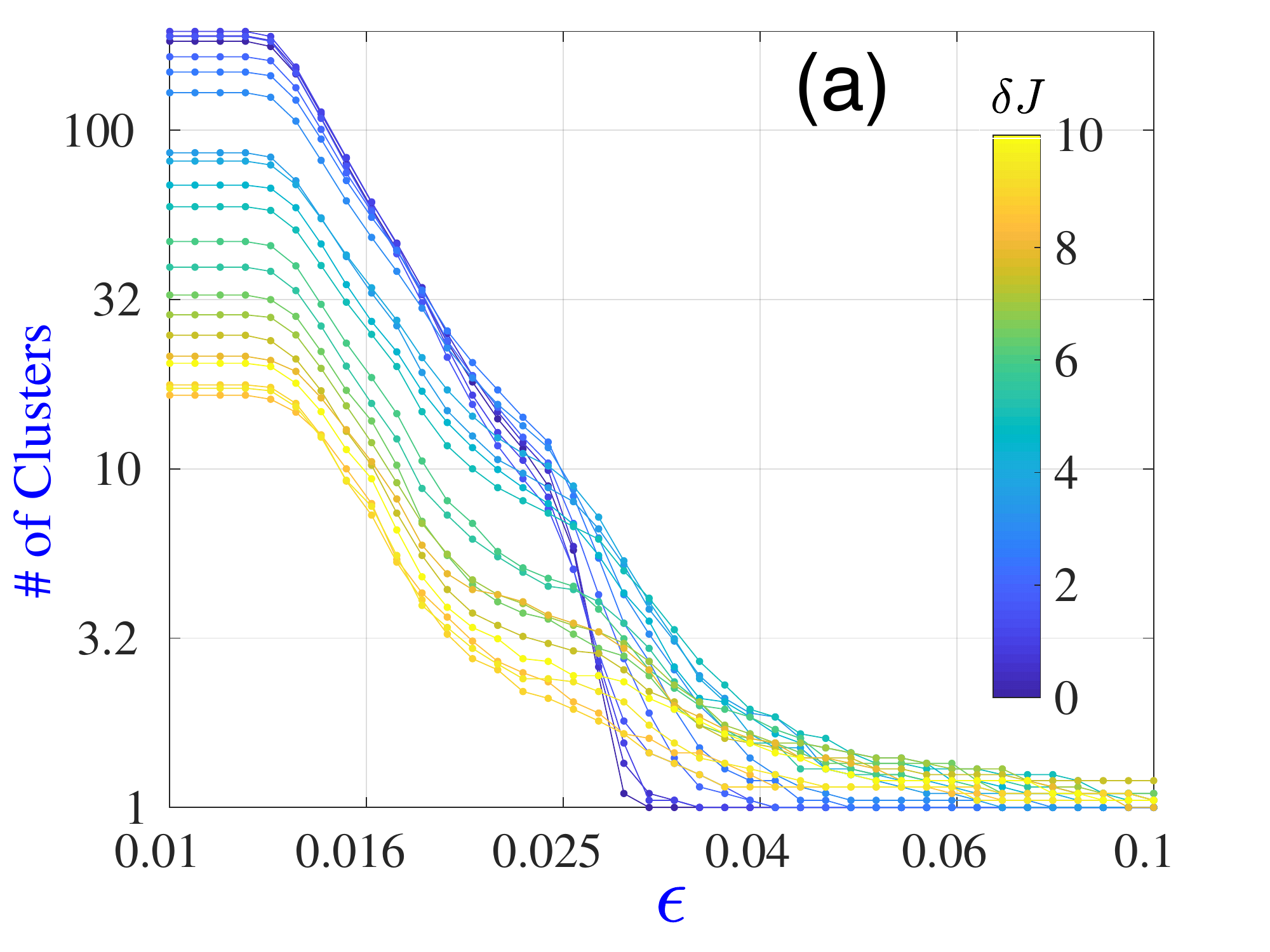}
	\includegraphics[width=0.49\linewidth]{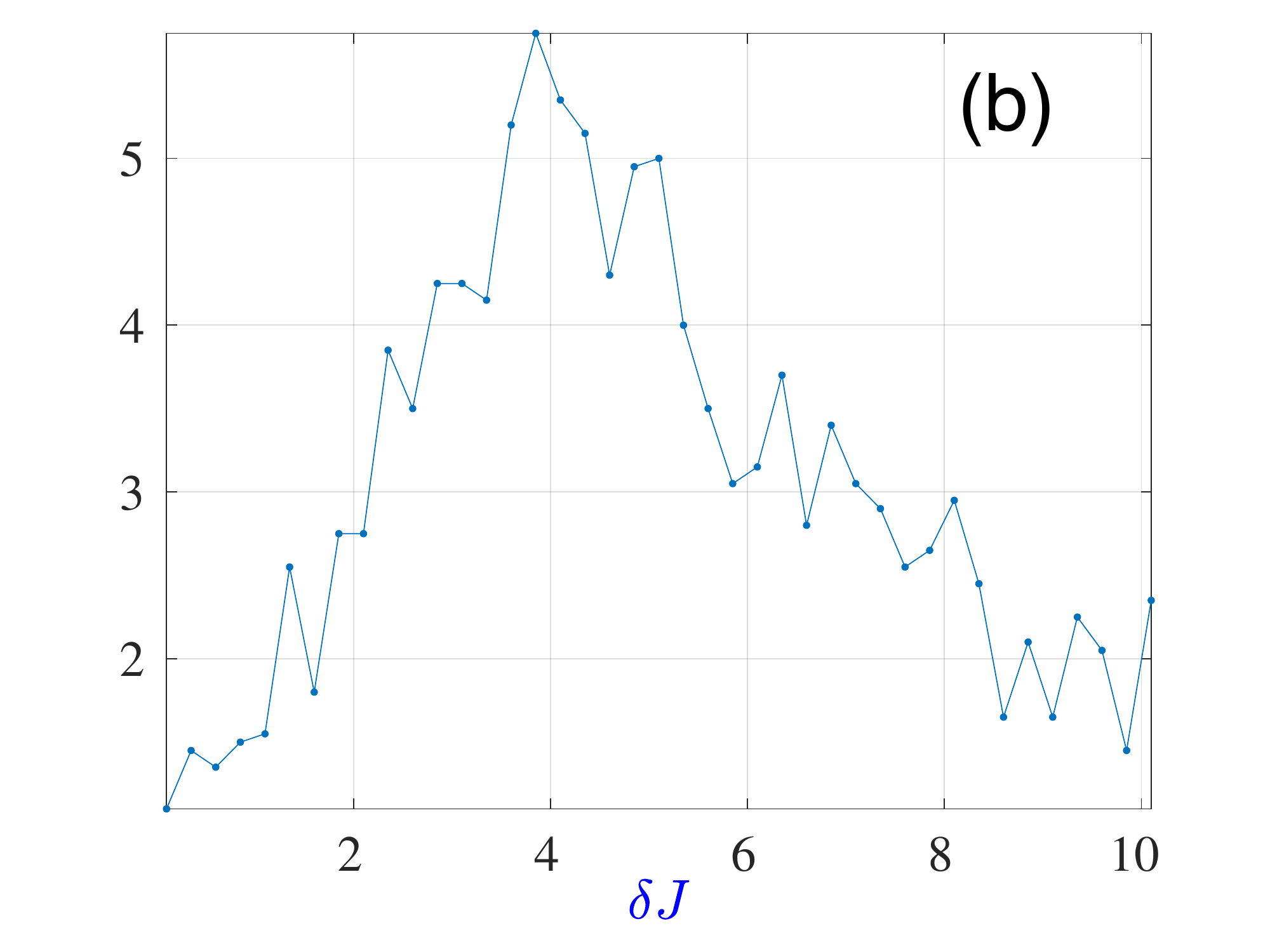}
	\caption{ Number of clusters learned by diffusion maps on the measurement samples of the long-time dynamical state of Eq.\,\eqref{H}, averaged over 50 disorder realizations. 500 samples are obtained for $N=12$ spins using exact diagonalization. (a) The number of clusters as a function of $\epsilon$. (b) For an intermediate value of $\epsilon$ ($\epsilon =0.029$), the number of clusters is peaked around the critical disorder strength $\delta J_c=3.81$.} 
	\label{J1_rand_fig}
\end{figure}

\bibliographystyle{apsrev4-1}
\bibliography{SM_bib}

\title{Supplementary Material for ``Unsupervised machine learning of quantum phase transitions using diffusion maps''}

\author{Alexander Lidiak}
\email{alidiak@mines.edu}
\affiliation{Department of Physics, Colorado School of Mines, Golden, Colorado 80401, USA}
\author{Zhexuan Gong}
\email{gong@mines.edu}
\affiliation{Department of Physics, Colorado School of Mines, Golden, Colorado 80401, USA}
\affiliation{National Institute of Standard and Technology, Boulder, Colorado 80305, USA}

\maketitle

In this supplementary material, we will show additional supporting results for how diffusion maps can learn incommensurate phases (section I), valence-bond solid phases (section II), and many-body localization (section III).

\section{Learning incommensurate phases}

\begin{figure}[htp]
	\centering
	\includegraphics[width=\textwidth]{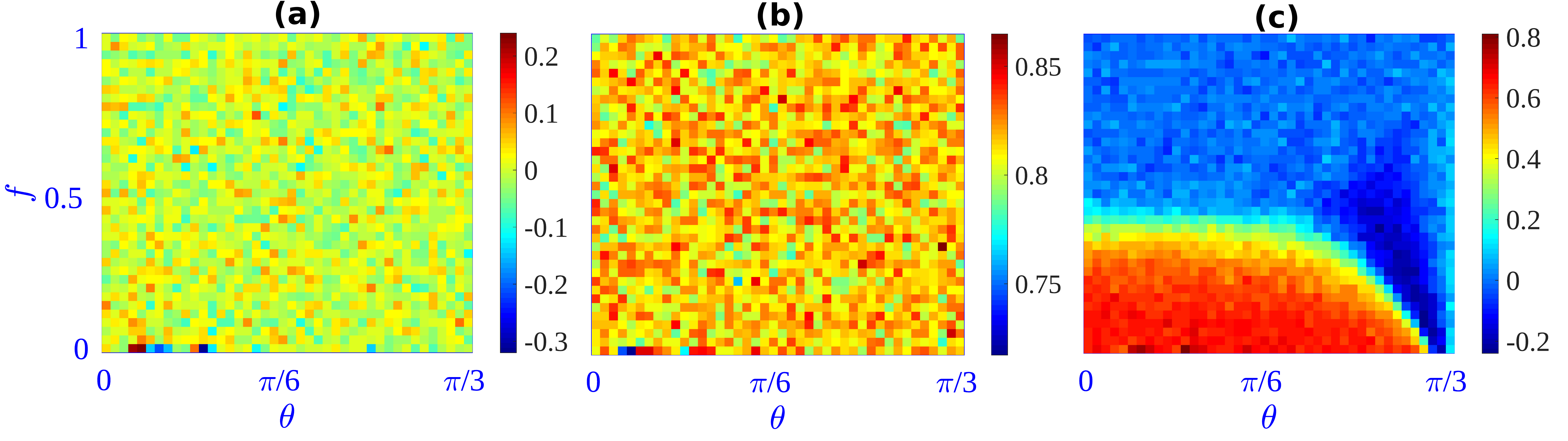}
	\caption{The expectation value of $\sum_i{\sigma_i/N}$ from $H_1$ (a), the variance of the same quantity (b), and the correlation between the middle and the end spins $\langle \sigma_1 \sigma_{N/2}\rangle$ (c) as a function of $f$ and $\theta$. Same as Fig.\,2 of the main text, $N=24$ and $500$ samples are used for each value of $f$ and $\theta$.}
	\label{clock3_observables}
\end{figure}

In this section, we first show that traditional data analysis methods in quantum simulation that focus on the expectation value, variance, or two-point correlations of local observables will not identify the incommensurate phase associated with $H_1$ (see main text). As seen from Fig.\,\ref{clock3_observables}, expectation values and variances of the $\sigma$ operators (averaged spatially) do not identify any phases. This is because the samples are obtained from the exact ground state of $H_1$ at a finite system size, thus spontaneous symmetry breaking does not happen. The two-point correlations of the $\sigma$ operators between the edge spin and the central spin is capable of revealing the ferromagnetic phase, but does not distinguish the incommensurate phase and the paramagnetic phase clearly.

Next, we show that principle component analysis (PCA) together with k-means clustering is also unable to learn the incommensurate phase and its  boundary. We perform PCA on the same collection of measurement samples used for the diffusion map in Fig\,2 of the main text and extract the projection of the sample set onto the first two principle components. We then apply a k-means clustering algorithm to associate each sample $\bm{X}_i$ with an index $L_i=1,2,\cdots,k$. In an attempt to identify all three phases within the samples, we manually set the number of clusters to $k=3$ (note that the diffusion map does not require such a priori knowledge of the number of distinct phases in the data). We then average $L_i$ for samples belonging to a particular set of Hamiltonian parameters ($f$, $\theta$), and obtain a phase diagram using this averaged index [see Fig.\,\ref{CCM_Kmeans}(a)]. While we can identify the ferromagnetic phase and its boundary, there is no clear identification of the paramagnetic or incommensurate phases, similar to the scenario with two-point correlations discussed above.

We have also used an auto-encoder included in MATLAB to perform nonlinear dimensionality reduction of the measurement data in substitution of PCA. The auto-encoder trains an artificial neural network to retain as much information of the sample data as possible with two latent variables onto which we then encode each measurement sample (similar to projecting onto the first two principle components obtained via PCA). Applying the same k-means clustering algorithm with $k=3$ leads to the phase diagram shown in Fig.\,\ref{CCM_Kmeans}(b), which again is unable to identify the incommensurate phase. This is because the incommensurate phase cannot be identified using a single linear or nonlinear function of the measured observables, as the spin configurations in this phase vary strongly with the system parameters ($\theta$ and $f$). The diffusion map method instead detects the change in the distribution of the measurement samples in configuration space which is often linked to a phase transition.

\begin{figure} [htp]
	\includegraphics[width=0.45\textwidth]{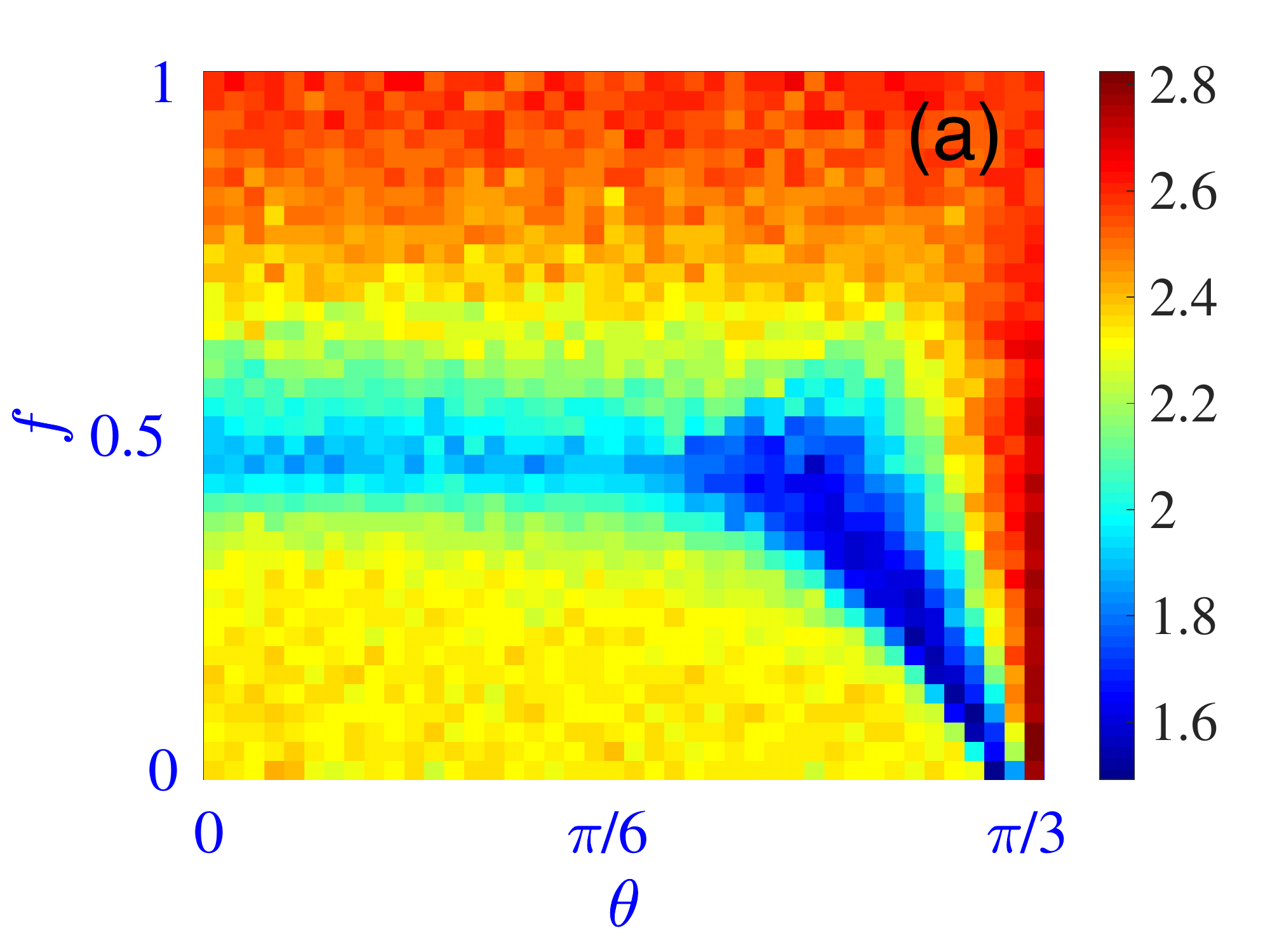} 
	\includegraphics[width=0.45\textwidth]{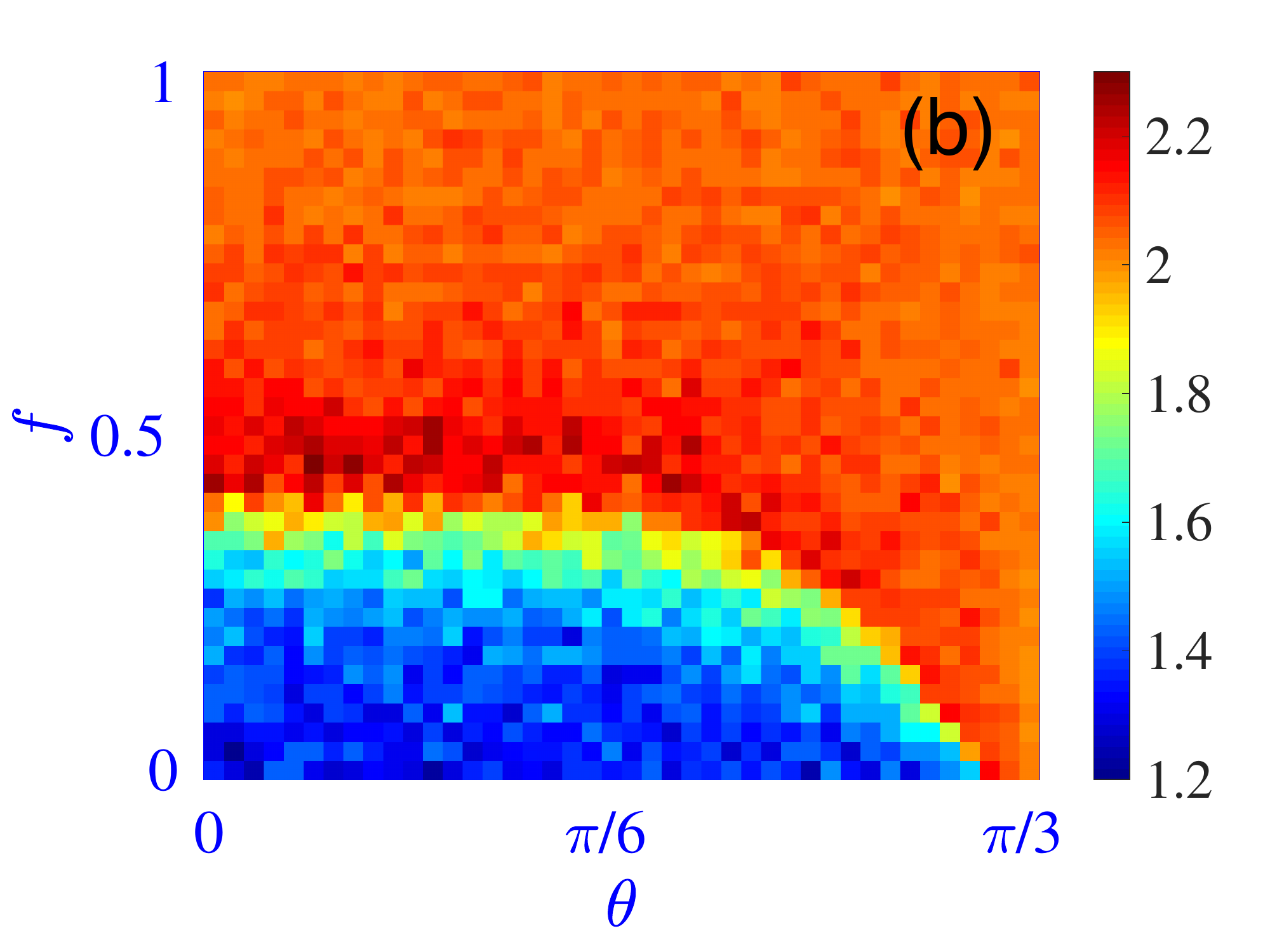}
	\caption{Phase diagrams generated by PCA (a) and auto-encoder (b) applied on the same measurement samples used in Fig.\,2 of the main text. The color represents the index attached by a $k=3$ k-means clustering algorithm on the projected/compressed measurement sample data, averaged over all measurement samples of the ground state of a particular Hamiltonian.}
	\label{CCM_Kmeans}
\end{figure}

The diffusion map method can also be applied to measurement samples of the $\tau$ operators in $H_1$. As shown in Fig.\,\ref{fig:Clock_diffmaps}(a), we see that the number of clusters identified by the diffusion map can also reveal all three phases and their boundaries. This is because if we obtain samples for the $\tau$ operator, then the paramagnetic phase will have the least number of clusters (approaching 1 deep in the paramagnetic phase), the ferromagnetic phase will have the most number of clusters and the incommensurate phase will have an intermediate number of clusters (as it is quasi long-ranged in the $\tau$ operator).  The phase diagram is thus nearly identical to Fig.\,2(a) of the main text with the color map flipped, the main difference being that the discrete $Z_3$ symmetry is identified in Fig.\,2 of the main text but not here, which is expected as the symmetry applies to the $\sigma$ operators only.

Finally, we show in Fig.\,\ref{fig:Clock_diffmaps}(b) a phase diagram of $H_1$ obtained using the half-chain entanglement entropy for the same system size and parameter range studied (similar to Ref.\,\cite{Zhuang2015}). One can see that the location and sharpness of the phase boundaries learned by the diffusion map method is comparable to that using the entanglement entropy.

\begin{figure}[htp]
    \includegraphics[width=0.9\textwidth]{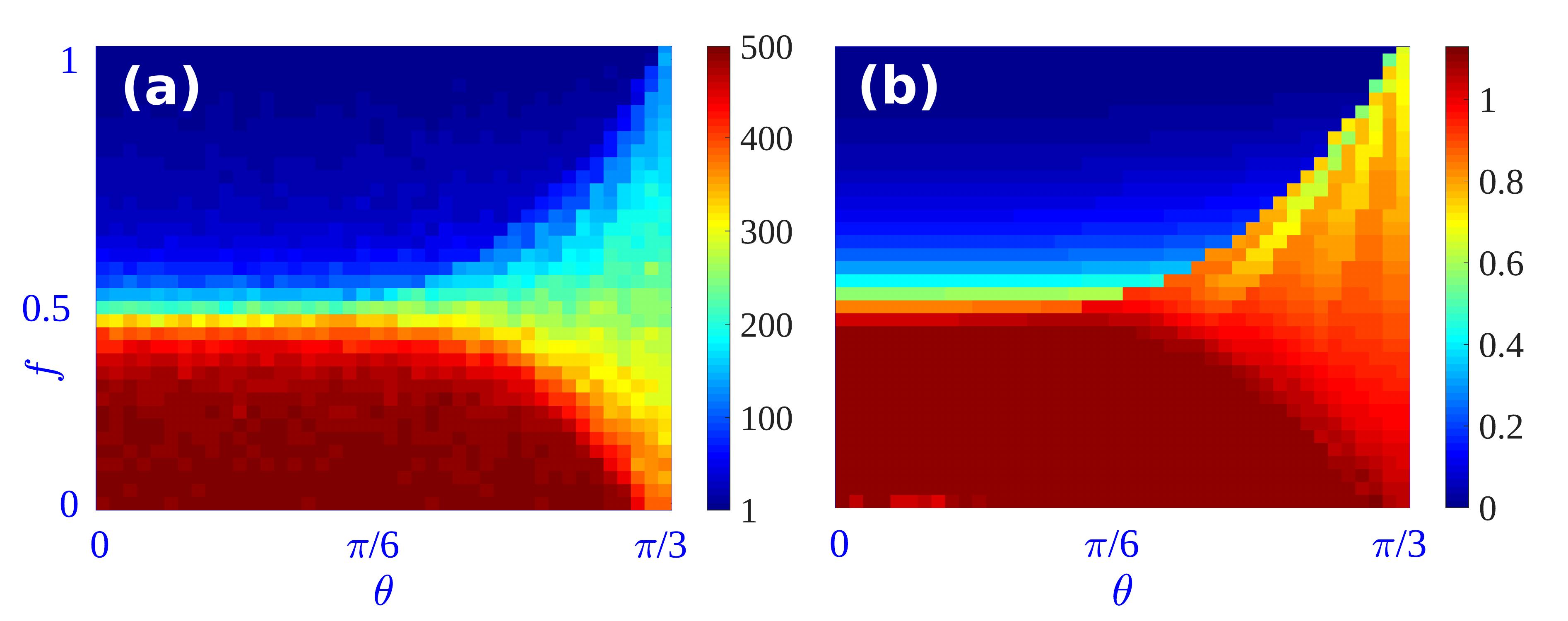}  
	\caption{(a) Ground-state phase diagram of $H_1$ (see main text) obtained by (a) performing diffusion maps with $\epsilon=0.025$, $\delta=10^{-2.5}$ on 500 measurement samples of the $\tau$ operators, and (b) calculating the half-system entanglement entropy.}
	\label{fig:Clock_diffmaps}
\end{figure}

\section{Learning valence-bond solid phases}
In this section, we will first show that for the $J_1$-$J_2$ model discussed in the main text ($H_2$), PCA and k-means clustering cannot detect the formation of valence-bond solids (VBS) or the spontaneous symmetry breaking at $J_2=0.5$. We first perform PCA on the same measurement samples used in Fig.\,3 of the main text. As shown in Fig.\,\ref{J1J2_kmeans}(a), the efficacy of dimensionality reduction is poor in this case, with many principle components contributing significantly to the variance of the data. Moreover, the first principle component only identifies an anti-ferromagnetic order, which is irrelevant for the VBS phase transition [Fig.\,\ref{J1J2_kmeans}(b)]. Keeping the first two principle components, we apply a k-means clustering algorithm with $k=2$ and plot the average index as a function of $J_2$ [Fig.\,\ref{J1J2_kmeans}(c)]. There is no clear signature of a phase transition happening at $J_2\approx 0.3$ and no indication of the spontaneous translational symmetry breaking at $J_2=0.5$. We have also used an auto-encoder in place of PCA and find it performs no better.

On the other hand, we show in Fig.\,\ref{J1J2_kmeans}(d) that the VBS phase transition can be identified using a complex order parameter known as the dimer correlation (see Eq.\,\eqref{dim_corr} and Ref.\,\cite{J1J2_Hikihara2001}). For the finite system size ($N=24$) studied here, the dimer correlation cannot locate the phase transition point exactly, and we find that the sharpness of the phase transition indicated by the dimer correlation is comparable to that obtained by the diffusion map method (see Fig.\,3(b) of the main text).

\begin{equation}
    C_{dim}^x(r)=\frac{1}{S^4} \langle S_{l_0-\frac{r}{2}}^x S_{l_0-\frac{r}{2}+1}^x (S_{l_0+\frac{r}{2}}^x S_{l_0+\frac{r}{2}+1}^x - S_{l_0+\frac{r}{2}+1}^x S_{l_0+\frac{r}{2}+2}^x) \rangle
    \label{dim_corr}
\end{equation}

\begin{figure}[h]
    \centering
		\includegraphics[width=1\textwidth]{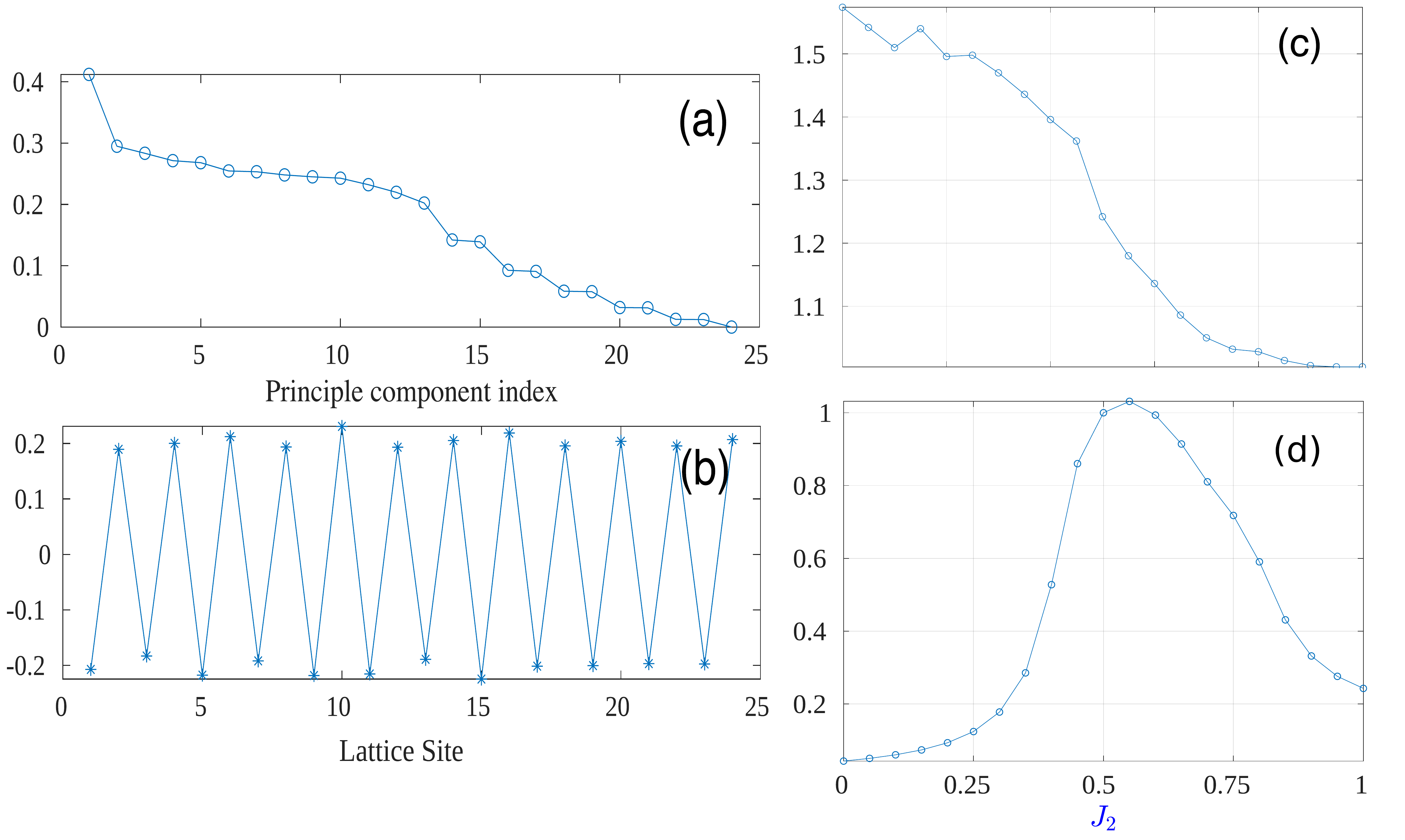}
	\caption{ (a) Principle values of PCA applied to the same measurement data used in Fig.\,3 of the main text. (b) Coefficients of the first principle component as a linear combination of each spin's magnetization, which indicates that the first principle component is an anti-ferromagnetic order parameter. (c) The average index assigned by k-means clustering with $k=2$ to the measurement samples projected onto the first two principle components, as a function of $J_2$. (d) The dimer correlation $C_{dim}^x(r)$ given by Eq.\,\eqref{dim_corr} as a function of $J_2$ calculated using the exact ground state of $H_2$. Here we set $l_0=13$ and $r=11$, consistent with the choice in Ref.\,\cite{J1J2_Hikihara2001}}.
	\label{J1J2_kmeans}
\end{figure}

Next, we will explain in detail the sharp drop in the number of clusters at $J_2=0.5$ identified by the diffusion map with an intermediate value of $\epsilon$. First, we point out that at $J_2=0.5$, we only obtain one of the two degenerate dimer states (one with $\bm{S}_i+\bm{S}_{i+1}=0$ for all odd $i$ and the other for all even $i$) as the ground state numerically, which is also expected in an actual experiment due to spontaneous symmetry breaking. A small deviation from $J_2=0.5$ will lift the degeneracy of the two dimer states and the ground state becomes approximately a superposition of the two dimer states, which we will call the `combined dimer state' below. Thus to see why the number of clusters identified by diffusion maps suddenly drops at $J_2=0.5$, we can compare the results of the diffusion map on a single dimer state with that of the combined dimer state. This comparison is shown in Fig.\,\ref{J1J2_Ndegen} and is very similar to how the $J_2=0.5$ curve compares to the $J_2$ close to $0.5$ curves in Fig.\,3(a) of the main text.

To understand why the number of clusters for the single dimer state decreases much more rapidly than the combined dimer state, let us start from the following analysis: if we get two random measurement samples from the combined dimer state, then there is $1/2$ probability that both samples are drawn from either one of the single dimer states (Case I), and $1/2$ probability that the two samples are drawn from two different single dimer states (Case II). The intuition is that the probability of finding two samples that are close to each other is much smaller in Case II than in Case I. For example, in Case I, for a given first sample, the chance of getting the second sample that has zero distance (i.e. identical) to the first sample is always $1/2^{N/2}$. But in Case II, the chance of getting a identical sample is $2/2^N$ because there are two identical samples that can be obtained from both the even and odd dimer states (which are the two perfect antiferromagnetic states), and each only appears with a probability of $1/2^{N}$. As a result, we can largely ignore the probability of finding two samples close to each other in case II. As we will show below, in the thermodynamic limit ($N\rightarrow\infty$), case II can be completely ignored except when we are considering two samples with exactly half of the spins in different directions. After ignoring case II, the probability of finding two samples with $k$ spins different for the combined dimer state, denoted by $P_c(k)$, is only half that of the single dimer state, denoted by $P_s(k)$, in the large $N$ limit.

The above analysis can be made precise mathematically. We find that $P_s(k)=\binom{\frac{N}{2}}{\frac{k}{2}}/2^{\frac{N}{2}}$ and $P_c(k)=[2^{\frac{N}{2}-1} \binom{\frac{N}{2}}{\frac{k}{2}}+\binom{N}{k}]/2^{N}$. We have plotted the ratio $P_s(k)/P_c(k)$ in Fig.\,\ref{J1J2_Ndegen}(b) for $N=24$ and $N=1000$. In the $N\rightarrow\infty$ limit, one can show analytically that $P_s(k)/P_c(k)=2$ for $k\ne N/2$ and $P_s(k)/P_c(k)=2/(1+\sqrt{2})\approx 0.83$ for $k=N/2$. Note that both $P_s(k)$ and $P_c(k)$ are symmetric around $k=N/2$.

Because the probability of finding samples for most distances in the single dimer state is two (or close to two for finite $N$) times larger than that in the combined dimer state, the number of clusters for the single dimer state will decrease with the cluster radius (proportional to $\epsilon$) at twice the rate of that for the combined dimer state. This twice as fast decay is what we observe in Fig.\,\ref{J1J2_Ndegen}(a) as well as Fig.\,3 of the main text which exhibits similar physics.

\begin{figure}
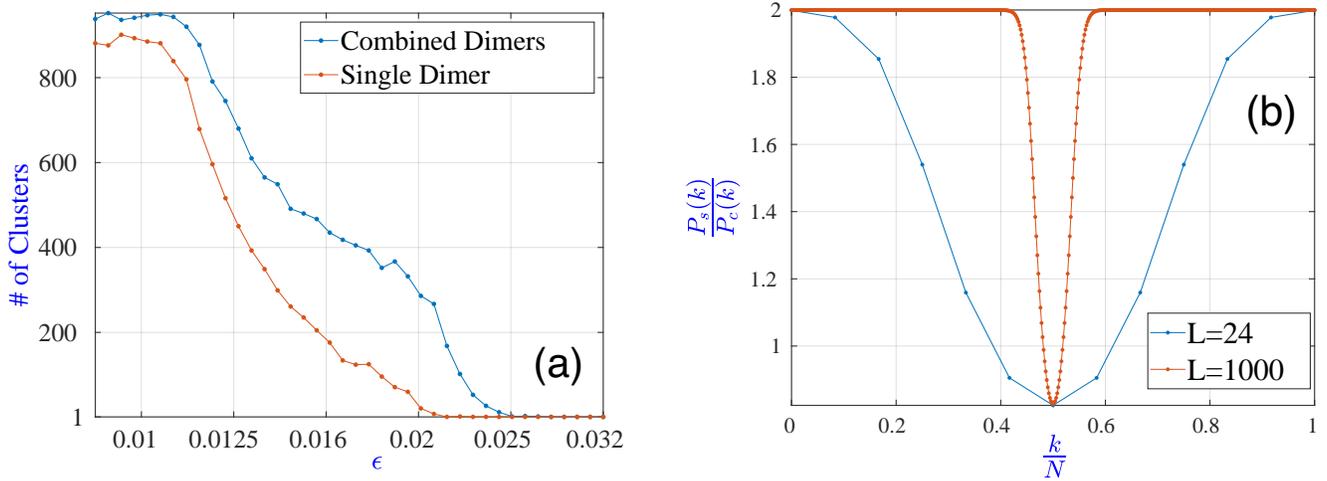


        \includegraphics[width=0.49\textwidth]{figures/supp_fig4a.pdf}\hfill
	    \includegraphics[width=0.5\textwidth]{figures/supp_fig4b.pdf}
	\caption{(a) The number of clusters identified by the diffusion map as a function of $\epsilon$ used on samples drawn from the single dimer state versus the combined dimer state with $N=24$. (b) The ratio $P_s(k)/P_c(k)$ of the probability of finding two samples with a difference of $k$ spins in the single dimer state to that in the combined dimer state.}
	\label{J1J2_Ndegen}
\end{figure}

\section{Learning many-body localization}

We will first explain why we use a different kernel in the diffusion map for learning many-body localization. In the many-body localized phase, the measurement samples will be clustered around the point of the initial state, while in the thermal phase, the measurement samples will be scattered across the configuration space. This leads to a large variation in the density of samples in configuration space, and the original Gaussian kernel (see main text) results in the number of clusters always decaying with increasing disorder strength for all values of $\epsilon$. This obscures the physical picture in that when the system goes from the thermal to the MBL phase, samples of the long-time dynamical states will start to form many small clusters due to the interplay of ergodicity and localization. To address this issue, we adopt a different kernel in the diffusion map that is widely used in data science on data sets with high density variations \cite{Coifman2006}, corresponding to an extra normalization of the Gaussian kernel (see the $\alpha=1$ case in Refs.\,\cite{Coifman2006,Rodriguez-Nieva2019}, i.e. we will use $K^{\prime}_{ij}=K_{ij}/(\sum_k K_{ik} \sum_k K_{kj})$ as the kernel in the place of $K_{ij}$). This extra normalization performed on $K_{ij}$ eliminates the sample density dependence of the diffusion map when extracting the structure of the samples in configuration space \cite{Coifman2006}. In the context of the thermal-to-MBL phase transition, we see that if the samples are drawn from a disordered state (high $h$ in $H_3$), then most of the samples are close to each other, with only a small amount of samples far away from the rest. Using the kernel given by $K^{\prime}_{ij}$, the transition probabilities between samples that are closely packed (far apart) will be reduced (increased), effectively evening out the density of samples in the configuration space. We note that for models without a big variation in sample density (e.g. those generated from the $J_1$-$J_2$ and chiral clock models we studied), using either $K_{ij}$ or $K^{\prime}_{ij}$ makes little difference.

Second, we show how diffusion maps can learn the many-body localization phase transition of a different model than the one studied in the main text. This model is a disordered spin-1/2 transverse field Ising chain with next-nearest neighbor interaction, with Hamiltonian:
\begin{equation}
    H_4= - \sum^{N-1}_{i=1} J_i \sigma_i^z \sigma^z_{i+1} +J_2 \sum^{N-2}_{i=1} \sigma_i^z \sigma^z_{i+2}+h \sum^N_{i=1} \sigma^x_i. \label{H}
\end{equation}
Here the nearest neighbor Ising couplings are disordered as $J_i=J+\delta J_i$ with $\delta J_i$ drawn from a uniform random distribution $[-\delta J,\delta J]$. According to Ref.\,\cite{J1rand_MBL}, this Hamiltonian undergoes a thermal to MBL phase transition near $\delta J_c=3.81 \pm 0.04$ (depending also on the energy density of the initial state) when $J=1$ and $\frac{h}{2}=J_2=0.3$. In Fig.\,\ref{J1_rand_fig}(a), we show the number of clusters of the diffusion map applied to the measurement samples of $\{\sigma_i^z\}$ drawn from the long-time dynamical state of an initial antiferromagnetic spin state as a function of $\epsilon$. We find that for intermediate values of $\epsilon$, the number of clusters shows a peak around $\delta J_c$ [Fig.\,\ref{J1_rand_fig}(b)]. This provides further evidence that diffusion maps are able to provide signatures of thermal to MBL phase transitions in general.

\begin{figure}[h]
	\includegraphics[width=0.49\linewidth]{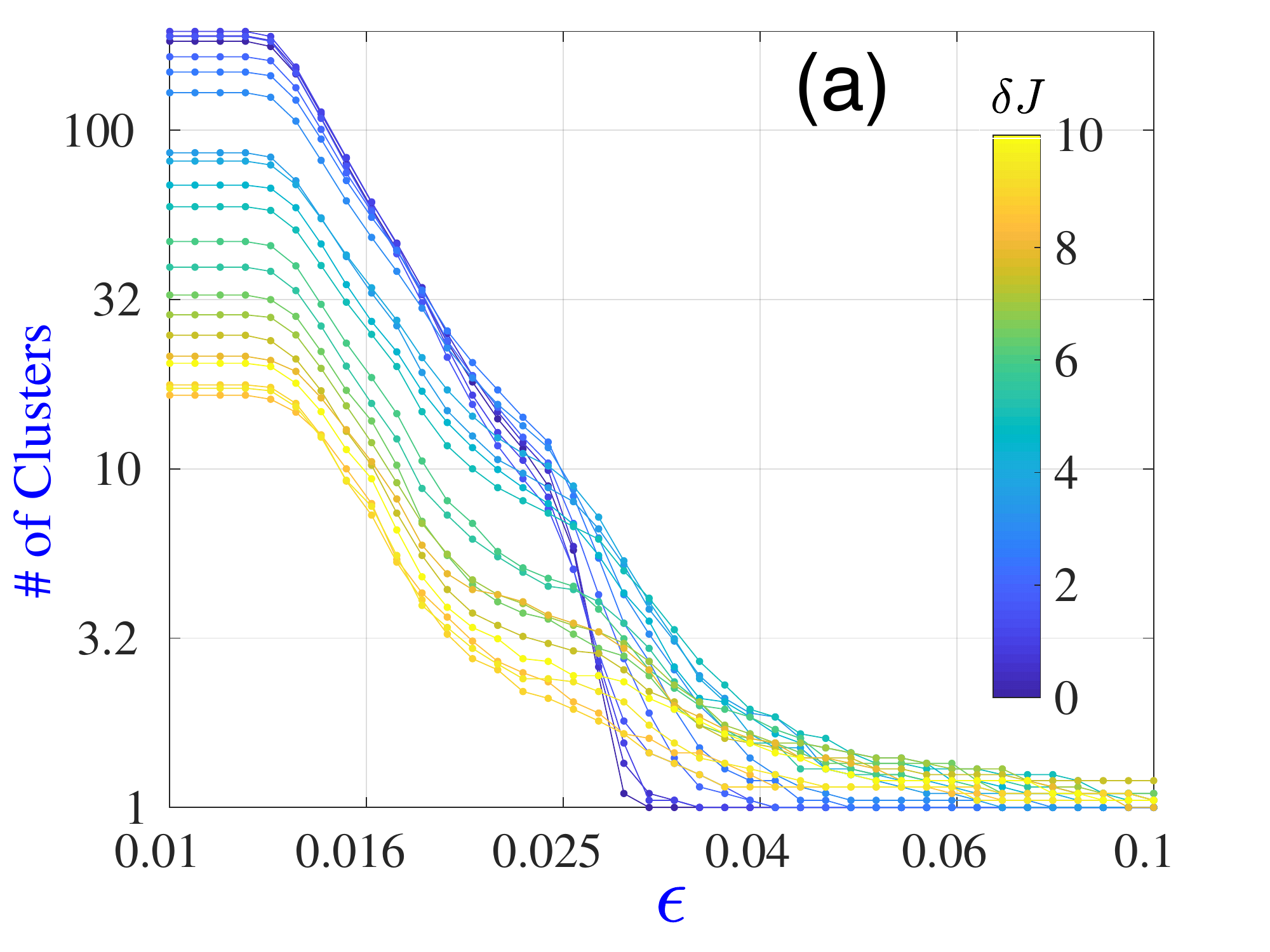}
	\includegraphics[width=0.49\linewidth]{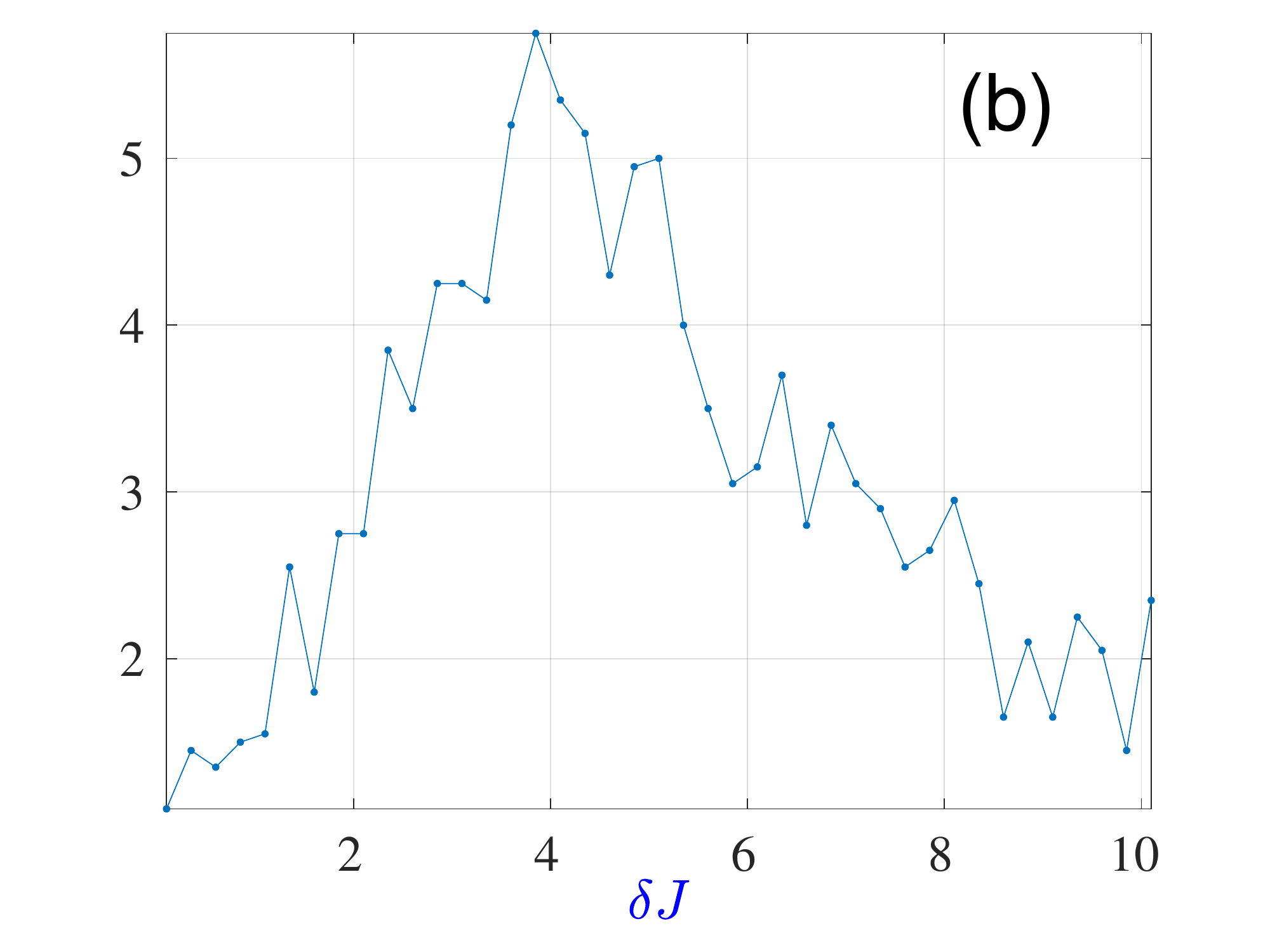}
	\caption{ Number of clusters learned by diffusion maps on the measurement samples of the long-time dynamical state of Eq.\,\eqref{H}, averaged over 50 disorder realizations. 500 samples are obtained for $N=12$ spins using exact diagonalization. (a) The number of clusters as a function of $\epsilon$. (b) For an intermediate value of $\epsilon$ ($\epsilon =0.029$), the number of clusters is peaked around the critical disorder strength $\delta J_c=3.81$.} 
	\label{J1_rand_fig}
\end{figure}

\bibliographystyle{apsrev4-1}
\bibliography{SM_bib}